\documentclass[12pt]{article}
\usepackage{fullpage}
\usepackage{epsf}
\usepackage{epsfig}
\usepackage{amsthm}
\usepackage{amsfonts}
\usepackage{color}
\usepackage{amsmath}
\usepackage{setspace}
\usepackage{natbib}
\usepackage{url}
\usepackage{algorithmic}
\usepackage[boxed]{algorithm}

\begin{document}

\newcommand{\bm}[1]{\mbox{\boldmath $#1$}}
\newcommand{\mb}[1]{\mathbf{#1}}
\newcommand{\bE}[0]{\mathbb{E}}
\newcommand{\bP}[0]{\mathbb{P}}
\newcommand{\ve}[0]{\varepsilon}
\newcommand{\Var}[0]{\mathbb{V}\mathrm{ar}}
\newcommand{\Corr}[0]{\mathbb{C}\mathrm{orr}}
\newcommand{\Cov}[0]{\mathbb{C}\mathrm{ov}}
\newcommand{\mN}[0]{\mathcal{N}}
\newcommand{\iidsim}[0]{\stackrel{\mathrm{iid}}{\sim}}
\newcommand{\NA}[0]{{\tt NA}}
\newcommand{\argmax}{\operatornamewithlimits{argmax}}

\title{\vspace{-1cm}Speeding up neighborhood search in\\local Gaussian process prediction}
\author{Robert B.~Gramacy\\
  Booth School of Business\\
  The University of Chicago\\
    {\tt rbgramacy@chicagobooth.edu}
  \and
  Benjamin Haaland\\
  Duke-NUS Graduate Medical School\\
  National University of Singapore\\
  {\tt ben.haaland@isye.gatech.edu}
}
 \date{}

\maketitle

\vspace{-0.2cm}
\begin{abstract}
Recent implementations of local approximate Gaussian process models have
pushed computational boundaries for non-linear, non-parametric prediction
problems, particularly when deployed as emulators for computer experiments.
Their flavor of spatially independent computation accommodates massive
parallelization, meaning that they can handle designs two or more orders of
magnitude larger than previously.  However, accomplishing that feat can still
require massive computational horsepower.  Here we aim to ease that burden. We
study how predictive variance is reduced as local designs are built up for
prediction.  We then observe how the exhaustive and discrete nature of an
important search subroutine involved in building such local designs may be
overly conservative. Rather, we suggest that searching the space radially,
i.e., continuously along rays emanating from the predictive location of
interest, is a far thriftier alternative.  Our empirical work demonstrates
that ray-based search yields predictors with accuracy comparable to exhaustive
search, but in a fraction of the time---for many problems bringing a
supercomputer implementation back onto the desktop.

\end{abstract}

\section{Introduction}
\label{sec:intro}

Gaussian process (GP) regression is popular for response surface modeling
wherever surfaces are reasonably smooth, but where otherwise little is known
about the the input--output relationship. GP regression models are
particularly popular as emulators for computer experiments
\citep{sack:welc:mitc:wynn:1989,sant:will:notz:2003}, whose outputs tend to
exhibit both qualities. Moreover computer experiments are often deterministic,
and it turns out that GPs are one of a few flexible regression approaches
which can interpolate while also predicting accurately with appropriate
coverage out-of-sample. Unfortunately GP regression requires $O(N^3)$ dense
matrix decompositions, for $N$ input--output pairs, so implementations
struggle to keep up with today's growing pace of data collection. Modern
supercomputers make submitting thousands of jobs as easy as submitting one,
and therefore $N=27$ runs is no longer a prototypically-sized
computer experiment \citep{chen:loeppky:sacks:welch}.

Most modern desktops cannot decompose dense matrices as large as $10^4
\times 10^4$, due primarily to memory constraints.  They struggle to perform
 tens to hundreds of much smaller, $10^3 \times 10^3$ decompositions required
for numerical inference of unknown parameters, say by maximum likelihood, in
reasonable time. Although humbly small by comparison to other literatures,
like genetics or marketing, those limits define ``big data'' for
computer experiments: the canonical models cannot cope with the modern scale
of computer simulation.

A scramble is on for fast, approximate, alternatives
\citep[e.g.,][]{kaufman:etal:2012,sang:huang:2012,gramacy:apley:2014}, and a
common theme is sparsity, leading to fast matrix decompositions.  Some
approaches  increase data size capabilities by one-to-two orders of
magnitude. Yet even those inroads are at capacity. Practitioners
increasingly prefer much cruder alternatives, for example trees
\citep{pratola:etal:2013,gra:tadd:wild:2012,chipman:rajan:wang:2012}
which struggle to capture the smoothness
of most simulators.

Another approach is to match supercomputer simulation with supercomputer
emulation, e.g, using graphical processing units
  \citep[GPUs,][]{franey:ranjan:chipman:2012}, cluster, and
symmetric-multiprocessor computation, and even all three together
\citep{paciorek:etal:2013}, for (dense matrix) GP regression.  This too has
led to orders of magnitude expansion in capability, but may miss the point:
emulation is meant to {\em avoid} further computation.  Hybrid approximate GP
regression and big-computer resources have been combined to push the boundary
even farther \citep{eidsvik2013estimation}.
\citet{gramacy:niemi:weiss:2014} later showed how GPUs and/or thousands of CPUs, distributed
across a cluster, could be combined to handle designs as large as $N=10^6$ in
about an hour.

This paper makes three contributions to this literature, focusing on a
particular sparsity-inducing local GP approximation developed by
\citet{gramacy:apley:2014}.  First, we study a greedy search subroutine, applied
locally and independently for each element of a potentially vast predictive
grid. Each search identifies small local sub-designs by a variance reduction
heuristic---the main sparsity-inducing mechanism---and we identify how it
organically facilitates a desirable trade-off between local and (more) global
site selection. We then observe highly regular patterns in the variance
reduction surface searched in each iteration of the greedy scheme, motivating
our second contribution.  We propose swapping out an exhaustive discrete
search for a continuous one having far narrower scope: along rays emanating
from each predictive site. Our empirical work demonstrates that the new scheme
yields accurate predictions in time comparable to a GPU/cluster computing
implementation, yet only requires a modern desktop. Third, to acknowledge that
rays can be inefficient in highly anisotropic contexts, we illustrate how a
thrifty pre-scaling step from a crude global analysis leads to improved
out-of-sample performance. Finally, the software modifications required are
slight, and are provided in the updated {\tt laGP} package for {\sf R}
\citep{laGP,gramacy:2014}.

The remainder of the paper is outlined as follows.  In Section
\ref{sec:relwork} we survey the modern literature for fast approximate GP
regression with an emphasis on sparsity and local search.   Section
\ref{sec:explore} explores the structure of local designs, recommending the
 simple heuristic presented in Section \ref{sec:exploit}. In Section
\ref{sec:illus} we provide implementation details and illustrations on real
and simulated data from the recent literature. Section
\ref{sec:discuss} concludes with a brief discussion.

\section{Fast approximate Gaussian process regression}
\label{sec:relwork}

Here we review the basics of GP regression, emphasizing computational
limitations and remedies separately leveraging sparsity and distributed
computation.  That sets the stage for a modern re-casting of a localization
technique from the spatial statistics literature that is able to leverage both
sparsity and big computing paradigms.  Yet even that method has
inefficiencies, which motivates our contribution.

\subsection{Kriging and sparsity}
\label{sec:krig}

A Gaussian process (GP) is a prior over functions
\citep[see, e.g.,][]{stein:1999}, with finite dimensional distributions
defined parsimoniously by a mean and covariance, often paired with a error
model (also Gaussian) for noisy data.  However for regression applications, a
likelihood perspective provides a more expedient view of the salient
quantities. In a typical GP regression, $N$ data
pairs $D_N = (X_N, Y_N)$, comprised of an $N\times p$-dimensional design $X_N$
and an $N$-vector of scalar responses $Y_N$, is modeled as $Y_N| \theta \sim
\mN_N(f_\theta(X_N),
\Sigma_\theta(X_N))$, where $\theta$ are a small number of parameters that relate
the mean $f$ and covariance $\Sigma$ to covariates $X_N$.  Linear regression
is a special case where $f_\theta(X_N) = X_N \beta$ and $\Sigma_\theta(X_N) =
\tau^2 I_N$.  

In the non-linear case it is typical, especially for computer experiments
\citep[e.g.,][]{sant:will:notz:2003}, to have a zero mean and therefore move
all of the ``modeling'' into a correlation function $K_\theta(x, x')$ so that
$Y_N \sim \mN_N(0, \tau^2 K_N)$ where $K_N$ is a $N \times N$ positive
definite matrix comprised of pairwise evaluations $K_\theta(x_i, x_j)$ on the
rows of $X_N$.  Choice of $K_\theta(\cdot, \cdot)$ determines the decay of
spatial correlation throughout the input space, and thereby stationarity and
smoothness.  A common first choice is the so-called {\em isotropic Gaussian}:
$K_\theta(x, x') = \exp\{-\sum_{k=1}^p (x_k - x'_k)^2/\theta\}$, where
correlation decays very rapidly with lengthscale determined through $\theta$.
Since $K_\theta(x, x) = 1$ the resulting regression function is an
interpolator, which is appropriate for many deterministic computer
experiments.  For noisy data, or for more robust modeling (protecting against
numerical issues as well as inappropriate choice of covariance, e.g., assuming
stationarity) of deterministic computer experiments
\citep{gra:lee:2012}, a {\em nugget} can be added to $K_\theta$.  In this
paper we will keep it simple and use the isotropic Gaussian formulation, and
fix a small nugget for numerical stability.  However, when appropriate we will
explain why authors have made other choices, often for computational reasons.
All new methodology described herein can be generalized to any correlation
family that is differentiable in $\theta$.

GP regression is popular because  inference (for $\theta$)
is easy, and (out-of-sample) prediction is highly accurate and conditionally
(on $\theta$) analytic.  It is common to deploy a reference $\pi(\tau^2)
\propto 1/\tau^2$ prior \citep{berger:deiliveira:sanso:2001} and obtain a marginal
likelihood for $\theta$
\begin{equation}
p(Y_N|K_\theta(\cdot, \cdot)) = \frac{\Gamma[N/2]}{(2\pi)^{N/2}|K_N|^{1/2}} \times
\left(\frac{\psi_N}{2}\right)^{\!-\frac{N}{2}},
\label{eq:gpk}
\;\;\;\;\; \mbox{where} \;\;\;
\psi_N = Y_N^\top K_N^{-1} Y_N,
\end{equation}
which has analytic derivatives, leading to fast Newton-like schemes for
maximizing.  

The predictive distribution $p(y(x) | D_N, K_\theta(\cdot, \cdot))$, is
Student-$t$ with degrees of freedom $N$,
\begin{align} 
  \mbox{mean} && \mu(x|D_N, K_\theta(\cdot, \cdot)) &= k_N^\top(x)  K_N^{-1}Y_N,
\label{eq:predgp} \\ 
\mbox{and scale} && 
 \sigma^2(x|D_N, K(\cdot, \cdot)) &=   
\frac{\psi_N [K_\theta(x, x) - k_N^\top(x)K_N^{-1} k_N(x)]}{N},
\label{eq:preds2}
\end{align}
where $k_N(x)$ is the $N$-vector whose $i^{\mbox{\tiny th}}$
component is $K_\theta(x,x_i)$.  Using properties of the Student-$t$,
 the variance of $Y(x)$ is $V_N(x) \equiv \Var[Y(x)|D_N, K_\theta(\cdot, \cdot)] =
\sigma^2(x|D_N,K_\theta(\cdot, \cdot))\times N/(N - 2)$.  The
form of $V_N(x)$, being small/zero for $x$'s in $X_N$ and expanding out in a
``football shape'' away from the elements of $X_N$, has attractive uses in design:
high variance inputs represent sensible choices for new simulations
\citep{gra:lee:2009}.

The trouble with all this is $K_N^{-1}$ and $|K_N|$, appearing in several
instances in Eqs.~(\ref{eq:gpk}--\ref{eq:preds2}), and requiring $O(N^3)$
computation for decomposing dense matrices.  That limits data size to $N
\approx 1000$ in reasonable time---less than one hour for inference and
prediction on a commensurately sized $N$-predictive-grid, say---without
specialized hardware.
\citet{franey:ranjan:chipman:2012} show how graphical processing unit (GPU)
matrix decompositions can accommodate $N \approx 5000$ in similar time.
\citet{paciorek:etal:2013} add an order of magnitude, to $N
\approx 60000$ with a combination of cluster computing (with nearly 100 nodes,
16 cores each) and GPUs.

An alternative is to induce sparsity on $K_N$ and leverage fast sparse-matrix
libraries, or to avoid large matrices all together. \citet{kaufman:etal:2012}
use a compactly supported correlation (CSC) function, $K_\theta(\cdot,
\cdot)$, that controls the proportion of $K_N$ entries which are non-zero;
\citet{sang:huang:2012} provide a similar alternative.
\citet{snelson:ghahr:2006} work with a
smaller $K_N$ based on a reduced global design of {\em pseudo-inputs}, while
\citet{cressie:joh:2008} use a truncated basis expansion.
Like the parallel computing options above, leveraging sparsity extends
the dense-matrix alternatives by an order of magnitude.  For example,
\citet{kaufman:etal:2012} illustrate with an $N = 20000$ cosmology dataset.

However, there is a need for bigger capability. For example,
\citet{pratola:etal:2013} choose a thrifty sum of trees model  which
allowed for cluster-style (message passing interface; MPI) implementation to
perform inference for a $N=$~7-million sized design distributed over hundreds of
computing cores.  Until recently, such a large data set was well out of reach
of GP based methods, whether by sparse approximation or distributed
computation.

\subsection{Local search}
\label{sec:lagp}

In computer experiments, where emulation or surrogate modeling emphasizes
accurate prediction (\ref{eq:predgp}--\ref{eq:preds2}),
\citet{gramacy:apley:2014} showed how orders of magnitude faster inference and
prediction can be obtained by modernizing {\em local kriging}
from the spatial statistics literature
\citep[][pp.~131--134]{cressie:1993}.  Focusing on quickly obtaining accurate
predictive equations at a particular location, $x$, local kriging involves
choosing data subsets $D_n(x) \subseteq D_N$, based on $n \ll N$ whose
$X_n(x)$ values are close to $x$. This recognizes that data far from $x$ have
vanishingly small influence on prediction given the typical choices of
rapidly decaying correlation functions $K_\theta(\cdot, \cdot)$.  The
simplest choice is to fill $D_n(x)$ with $n$ {\em nearest neighbors} (NNs)
$X_n(x)$ to $x$ in $X_N$, along with responses $Y_n(x)$.

This is a sensible idea.  As long as $D_n$ is determined purely by
computational considerations, i.e., not by looking $Y_N$ values, the result is
a valid probability model for $Y(x)|D_N$ \citep{datta:etal:2014}.  For modest
$n$, prediction and inference is fast and accurate, and as $n$ gets large
predictors increasingly resemble their full-data analogues with a variance
$V_n(x)$ that is organically inflated (higher variance for smaller $n$)
relative to $V_N(x)$.  Recently, \citet{emory:2009} showed that the NN version
works well for a wide swath of common choices of $K_\theta(\cdot,
\cdot)$. However, there is documented scope for improvement.
\citet{vecchia:1988} and \cite{stein:chi:welty:2004} argue that the NN designs
are sub-optimal---it pays to have a local design with more spread in the input
space.  However, finding an optimal local design $D_n(x)$, under almost any
criteria, would involve a combinatorially huge search even for modest $n$ and
$N$.

\citet{gramacy:apley:2014} showed how a greedy iterative search for local
designs, starting with a small NN set $D_{n_0}(x)$ and building up to $D_n(x)$
by augmenting $D_j(x)$, for $j=n_0, \dots, n-1$ through a simple objective
criteria  leads to better predictions than NN.  Importantly, the greedy and NN
schemes can be shown to have the same computational order, $O(n^3)$, when an
efficient updating scheme is deployed for each $j \rightarrow j+1$. The idea
of building up designs iteratively  for faster calculations is not new
\citep[][]{haaland:qian:2012,gramacy:polson:2011}, however the focus has
previously been global.  \citeauthor{gramacy:apley:2014}'s local search
chooses an $x_{j+1}$ from the remaining set $X_N \setminus X_j(x)$ to maximally reduce
predictive variance at $x$.  The local designs contain
a mixture of NNs to $x$ and somewhat farther out ``satellite'' points, which
we explore further in Section
\ref{sec:explore}.

The resulting local predictors are at least as accurate as other sparse methods,
like CSC, but incur an order of magnitude lower computational cost. Since
calculations are independent for each predictive location, $x$, prediction
over a dense grid can be trivially parallelized to leverage multiple cores
on a single (e.g., desktop) machine.  However somewhat dissapointingly, 
\citeauthor{gramacy:apley:2014} also observed that the resulting greedy local
predictors are not more accurate per-unit-computational cost than local NN
predictors derived from an order of magnitude larger $n$.  In other words, the
search for $x_{j+1}$ over a potentially huge number of candidates $|X_N \setminus
X_j(x)| = N-j$, for $j=n_0,
\dots,n$ is expensive relative to decomposing a larger (but still small
compared to $N$) matrix for GP inference and prediction.

\citet{gramacy:niemi:weiss:2014} later recognized that that search,
structured to independently entertain thousands of $x_{j+1}$ under identical
criteria, is ideal for exporting to a GPU.  Depending on the size of the search,
improvements were 20--100 fold, leading to global improvements (over all
$j=n_0, \dots, n$) of 5-20x, substantially out-pacing the
accuracy-per-flop rate of a big-NN alternative. Going further, they showed how
a GPU/multi-CPU/cluster scheme could combine to accurately emulate at
one-million-sized designs in about an hour---a feat unmatched by other
GP-based methodology.

Further savings are obtained by recognizing that searching over all $N-j$
candidates is overkill, whether via GPU or otherwise, since many are very far
from $x$ and thus have almost no influence on prediction.  Searching over $N'
\ll N-j$ NNs, say $N'=1000$ when $N=1$-million yields substantial speedups, but
multi-node/multi-core resources would still be required in large data
contexts.  Importantly, one must take care not to choose $N'$ so small as to
preclude entertaining the very points, well outside the NN set, which lead to
improvements over NN. Considering the observed regularity of the greedy local
designs, illustrated in more detail below, there does nevertheless seem to be
potential for drastically shortcutting an exhaustive search without limiting
scope, thereby avoiding the need for specialized (GPU) or cluster-computing
resources.

\section{Exploring local influence on prediction}
\label{sec:explore}

\begin{figure}[ht!]
\centering
\includegraphics[scale=0.54,trim=5 25 0 30]{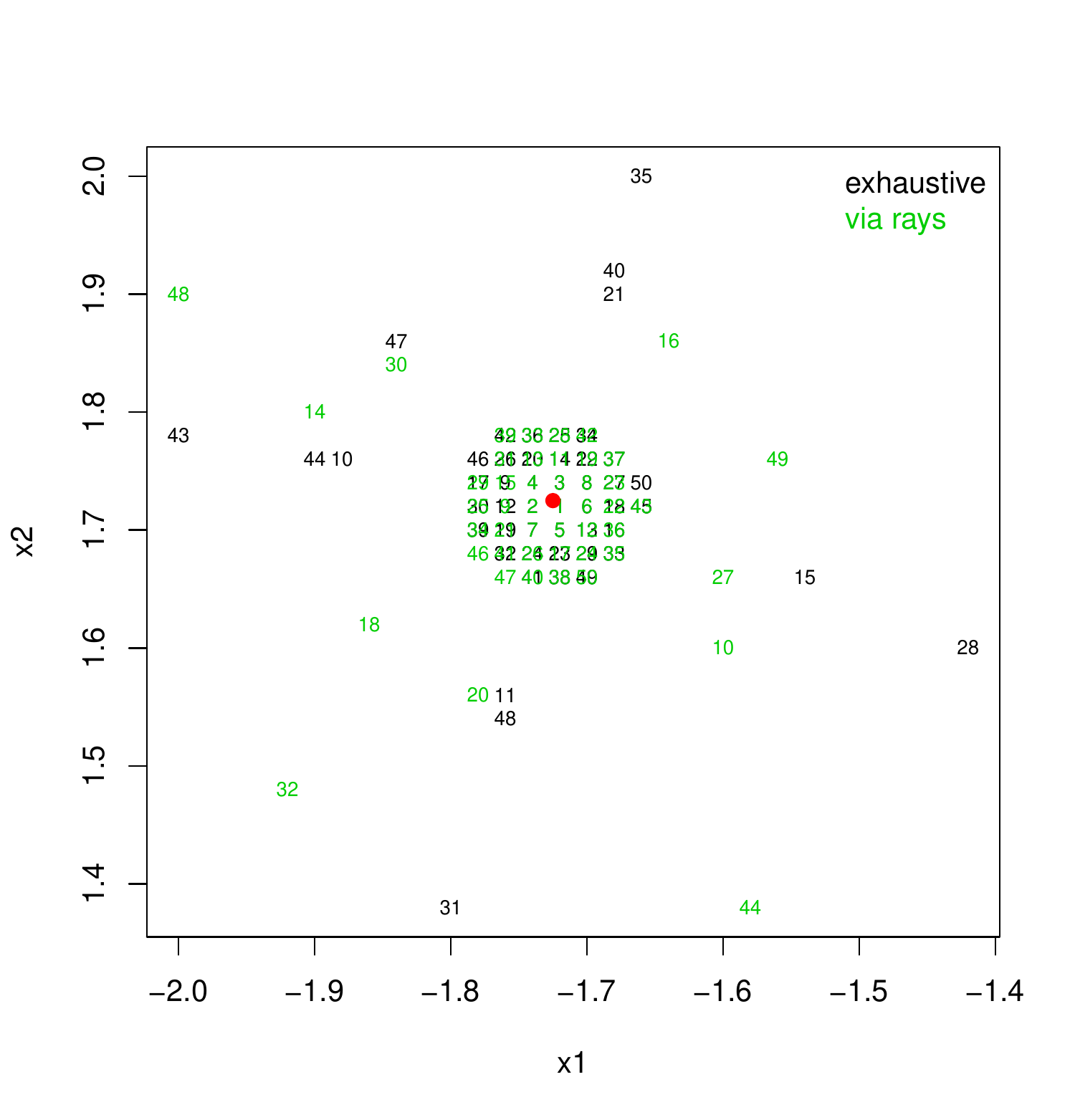} 
\caption{Example local designs constructed by a greedy application of
a reduction in variance criterion (\ref{eq:alc}).  The numbers indicate the
order in which each location is chosen.  The candidates form a regular $201
\times 201$ grid in $[-2,2]^2$; black numbers 
are from an exhaustive search [Section \ref{sec:explore}], whereas the green
ones from the search method proposed in Section \ref{sec:exploit}.}
\label{f:alc50}
\end{figure}

The numbers in Figure \ref{f:alc50} indicate local
designs $X_{n=50}(x)$ built up in a greedy fashion, successively choosing the
next location by a reduction in variance criteria, discussed in detail
shortly. Focus first on the black numbers, based on an exhaustive search. The
reference predictive location, $x=(-1.725,1.725)^\top$, is situated off of the
input design $X_{N=40401}$, which is a regular $201 \times 201$ grid in
$[-2,2]^2$. 
Observe that the local sub-design contains a roughly
$\frac{2}{3}\!:\!\frac{1}{3}$ split comprising of the nearest elements in
$X_N$ to $x$ and ones farther out. However even the farther-out elements are close
relative to the full scope of the grid in $[-2,2]^2$.  Observe that although
many NNs are chosen in early iterations, the farthest-out locations are not
exclusive to the final iterations of the search.  As early as $j=10$, a
far-out excursion was made, and as late as $j=50$ a NN was chosen.

On first encounter, it may be surprising that having farther-out design
elements helps reduce local predictive variance differentially more than
points much closer in.  One naturally wonders what trade-offs are
being made in the objective criteria for design, and to what extent they can be
attributed to a particular parameterization of $K_\theta(\cdot, \cdot)$. These
two aspects are explored below, leading to novel technical and empirical
observations, with an eye toward exploiting that trade-off for fast search in
Section \ref{sec:exploit}.

\subsection{From global to local design}
 
Designing $X_N$ to minimize predictive variance, averaged globally
over the input space $\mathcal{X}$, is a sensible  objective leading to
so-called $A$-optimal designs \citep[see, e.g.,][]{sant:will:notz:2003}.
However, minimizing over $N$ free coordinates in $p$-dimensional space is
combinatorially complex.
\citet{seo:etal:2000} showed that approximately $A$-optimal designs can
be obtained by proceeding sequentially, i.e., greedily: building up $X_N$
through choices of $x_{j+1}$ to augment $X_j$ to minimize variance globally.  In
particular, they considered choosing $x_{j+1}$ to minimize
\[
\Delta V_j(x_{j+1}) = \int_\mathcal{X} V_j(x) - V_j(x|
x_{j+1}) \, dx,
\]
where $V_j(x|x_{j+1})$ is the new variance after $x_{j+1}$ is added into
$X_j$, obtained by applying the predictive equations
(\ref{eq:predgp}--\ref{eq:preds2}) with a $(j+1)$-sized design
$[X_j; x_{j+1}^\top]$.  All quantities above, and below, depend implicitly on
$\theta$. Now, minimizing future variance is equivalent to maximizing a
quantity proportional to a {\em reduction} in variance:
\begin{equation} \label{eq:alc}
\int_\mathcal{X} k_j^\top(x) G_j(x_{j+1}) v_j(x_{j+1})
k_j(x) + 2k_j^\top(x) g_j(x_{j+1}) K(x_{j+1},x) + K(x_{j+1},x)^2 /
v_j(x_{j+1}) \; dx \vspace{-.5cm}
\end{equation}
\begin{align*}
\mbox{where } \quad\quad G_j(x') &\equiv g_j(x') g_j^\top(x'), \quad g_j(x') = K_j^{-1}
k_j(x')/v_j(x'), \\
v_j(x_{j+1}) &= K_\eta(x_{j+1},
x_{j+1}) - k_j^\top(x_{j+1}) K_j^{-1} k_j(x_{j+1}).
\end{align*}
The designs that result are
difficult to distinguish from other space-filling designs like maximin, maximum
entropy, $D$- or $A$-optimal.  The advantage is that greedy selection avoids a
combinatorially huge search.

\citet{gramacy:apley:2014} argued that the {\em integrand} in (\ref{eq:alc})
is a sensible heuristic for local design.  It tabulates a quantity
proportional to reduction in variance at $x$, obtained by choosing to add
$x_{j+1}$ into the design, which is the dominating component in an estimate of
the future mean-squared prediction error (MSPE) at that location.  When
applied sequentially to build up $X_n(x)$ via $X_{j}(x)$ and $x_{j+1}$ for
$j=n_0,
\dots, n-1$, the result is again an approximate a
solution to a combinatorially huge search. However, the structure of the
resulting local designs $X_n(x)$, with near as well as far points, is
counterintuitive [see Figure
\ref{f:alc50}]. The typical rapidly decaying
$K_\theta(\cdot, \cdot)$ should substantially devalue locations far from $x$.
Therefore, considering two potential locations, an $x_{j+1}$ close to $x$  and
$x_{j+1}'$ farther away, one wonders: how could the latter choice, $x_{j+1}'$ with
$y_{j+1}'$-value modeled as vastly less correlated with $y(x)$ than
$y_{j+1}$ via $x_{j+1}$, be preferred over the closer $x_{j+1}$ choice?

The answer is remarkably simple, and has little to do with the form of
$K_\theta(\cdot,\cdot)$. The integrand in Eq.~(\ref{eq:alc}) looks quadratic
in $K_\theta(x_{j+1}, x)$, the only part of the expression which measures a
``distance'', in terms of correlation $K_\theta$, between the reference
predictive location $x$ and the potential new local design location $x_{j+1}$.
That would seem to suggest maximizing the criteria involves maximizing
$K_\theta(x_{j+1}, x)$, i.e., choosing $x_{j+1}$ as close as possible to $x$.
But that's not the only part of the expression which involves $x_{j+1}$.
Observe that the integrand  is also quadratic in $g_j(x_{j+1})$, a vector
measuring ``inverse distance'', via $K_j^{-1}$, between $x_{j+1}$ and the
current local design $X_j(x)$. So there are two competing aims in the
criteria: minimize ``distance'' to $x$ while maximizing ``distance'' (or
minimizing ``inverse distance'') from the existing design.   Upon further
reflection, that tradeoff makes sense. The value of a potential new data
element $(x_{j+1}, y_{j+1})$ depends not just on its proximity to $x$, but
also on how potentially different that information is from where we already have
(lots of) it, at $X_j(x)$.

This observation also provides insight into the nature of global $A$-optimal
designs. We now recognize that the integral in Eq.~(\ref{eq:alc}), often
thought to be essential to obtain space-filling behavior in the resulting
design, is but one contributing aspect.  Assuming that potential design sites
for $X_N$ or $X_n(x) \subset X_N$ are limited, say to a pre-defined mesh of
values,\footnote{Note that this is always the case when working with a finite
precision computer implementation.} and that it is not possible to observe the
true output $y(x)$ at locations $x$ where you are trying to
predict,\footnote{That would cause the integrand in (\ref{eq:alc}) to be
minimized trivially, and obliterate the need for emulation.}  the GP predictor
prefers design sites which are somewhat spread out relative to where it will
be used to predict, globally or locally, regardless of the choice of
$K_\theta(\cdot, \cdot)$.

\subsection{Ribbons and rings}
\label{sec:rr}

Having established it is possible for the criteria to prefer new $x_{j+1}$
farther from $x$, being repelled by $X_j(x)$ already nearby, it is natural to
wonder about the extent of that trade-off in particular examples: of iteration
$j$, choice of parameters ($\theta$), and operational considerations such as
how searches are initialized (a number $n_0$ of NNs) and any numerical
considerations (i.e., nugget $\eta$).  Here we restrict $X_N$ to the $201
\times 201$ grid coinciding with the example in the previous subsection.
\begin{figure}[ht!]
\centering
\vspace{-0.14cm}
\begin{tabular}{c|c|c|c}
{\scriptsize $n_0 = 6$, $\theta=0.5$, $\eta=10^{-10}$} & 
{\scriptsize $n_0 = 6$, $\theta=0.5$, $\eta=10^{-3}$} &
{\scriptsize $n_0 = 6$, $\theta=0.1$, $\eta=10^{-3}$} & 
{\scriptsize $n_0 = 1$, $\theta=0.5$, $\eta=10^{-10}$} \\
\includegraphics[scale=0.225,trim=35 0 15 0]{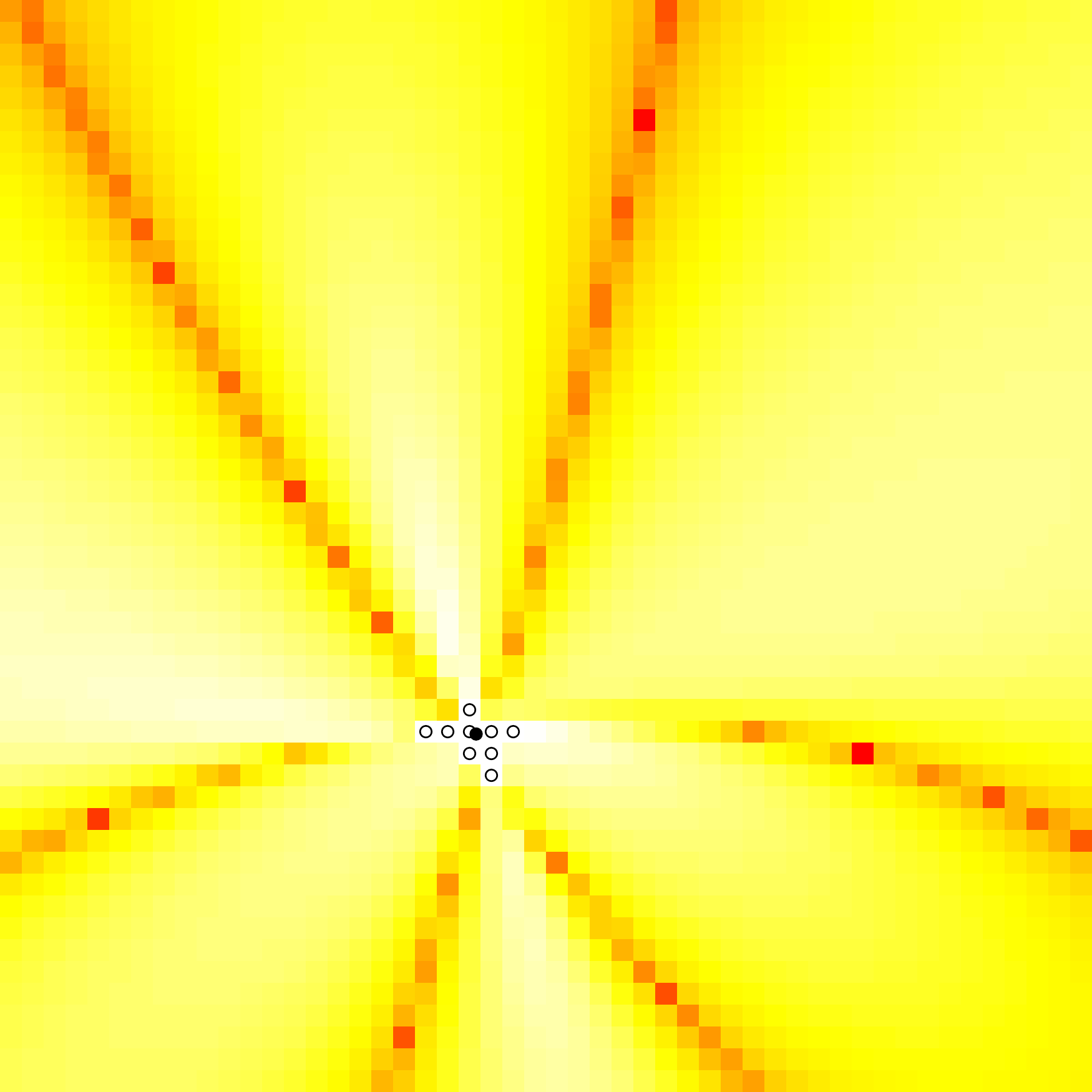} & 
\includegraphics[scale=0.225,trim=15 0 15 0]{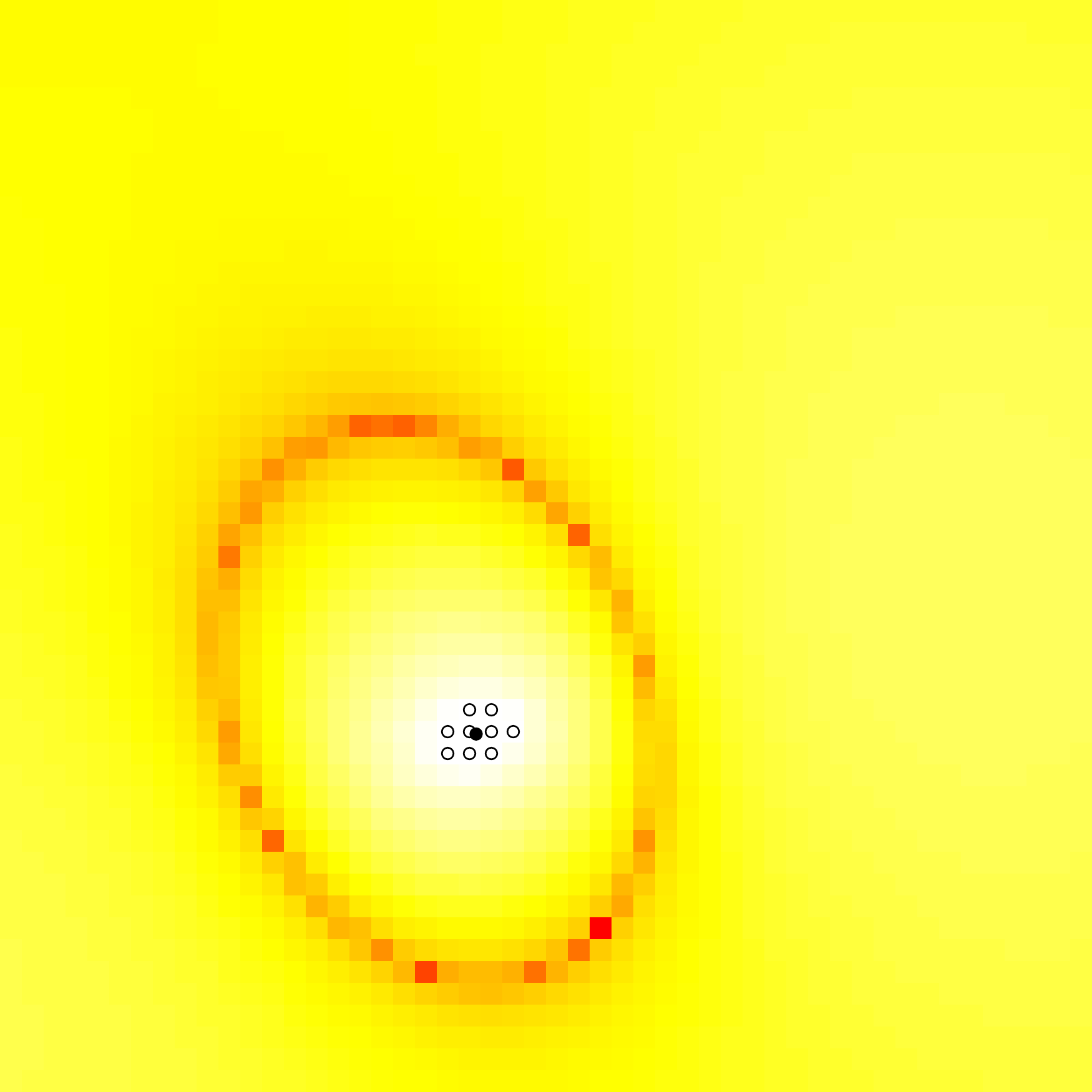} &
\includegraphics[scale=0.225,trim=15 0 15 0]{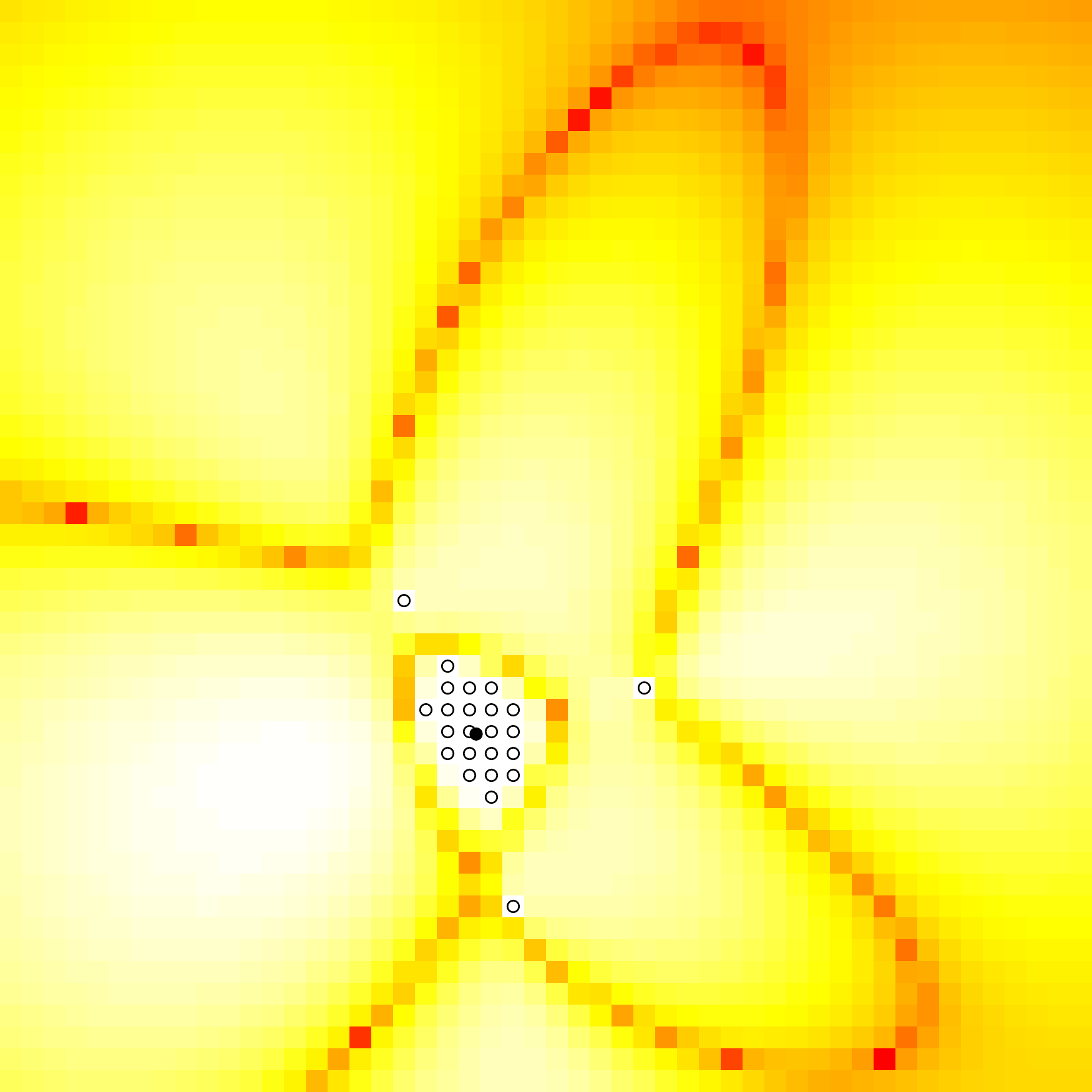} & 
\includegraphics[scale=0.225,trim=15 0 0 0]{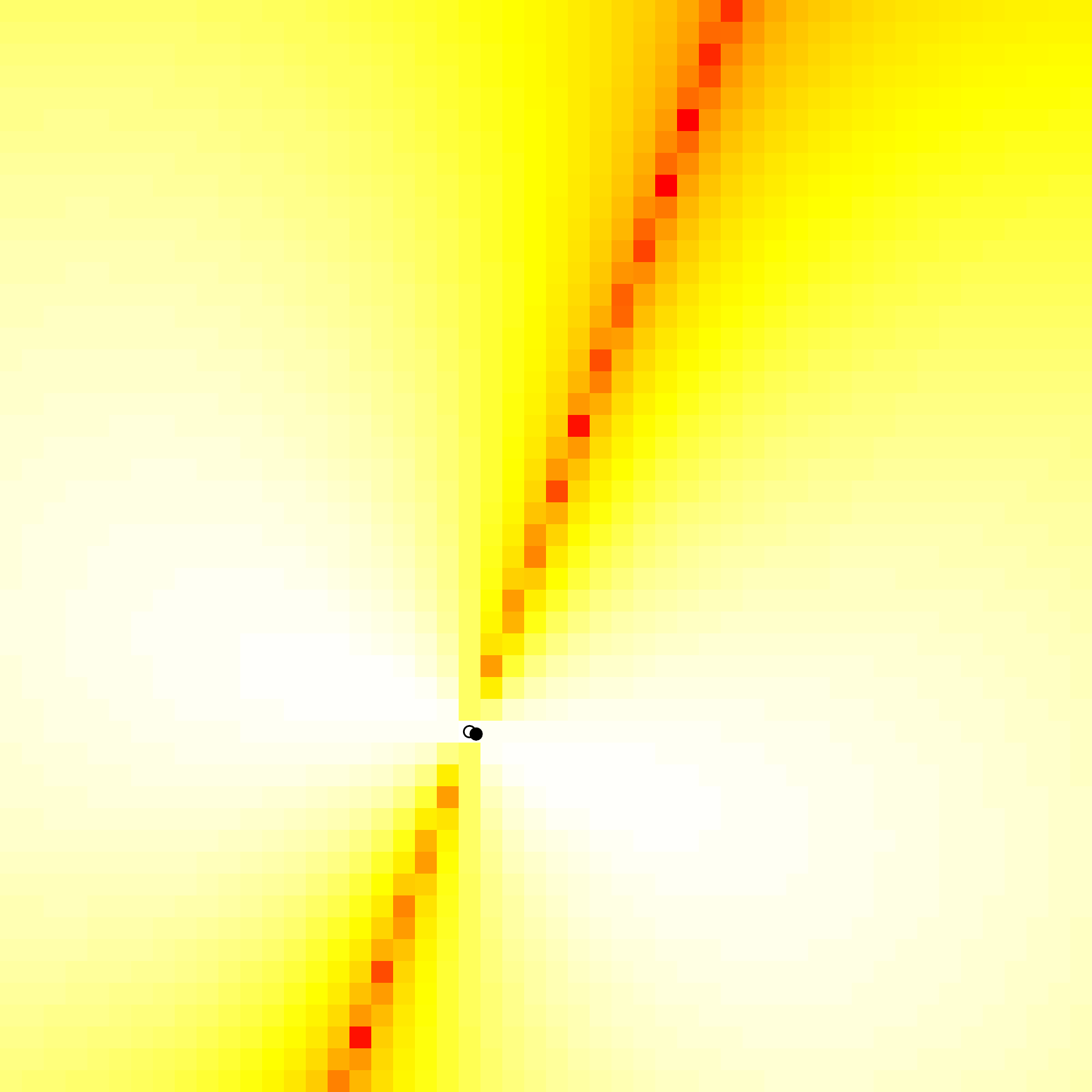}
\\
\includegraphics[scale=0.225,trim=35 0 15 0]{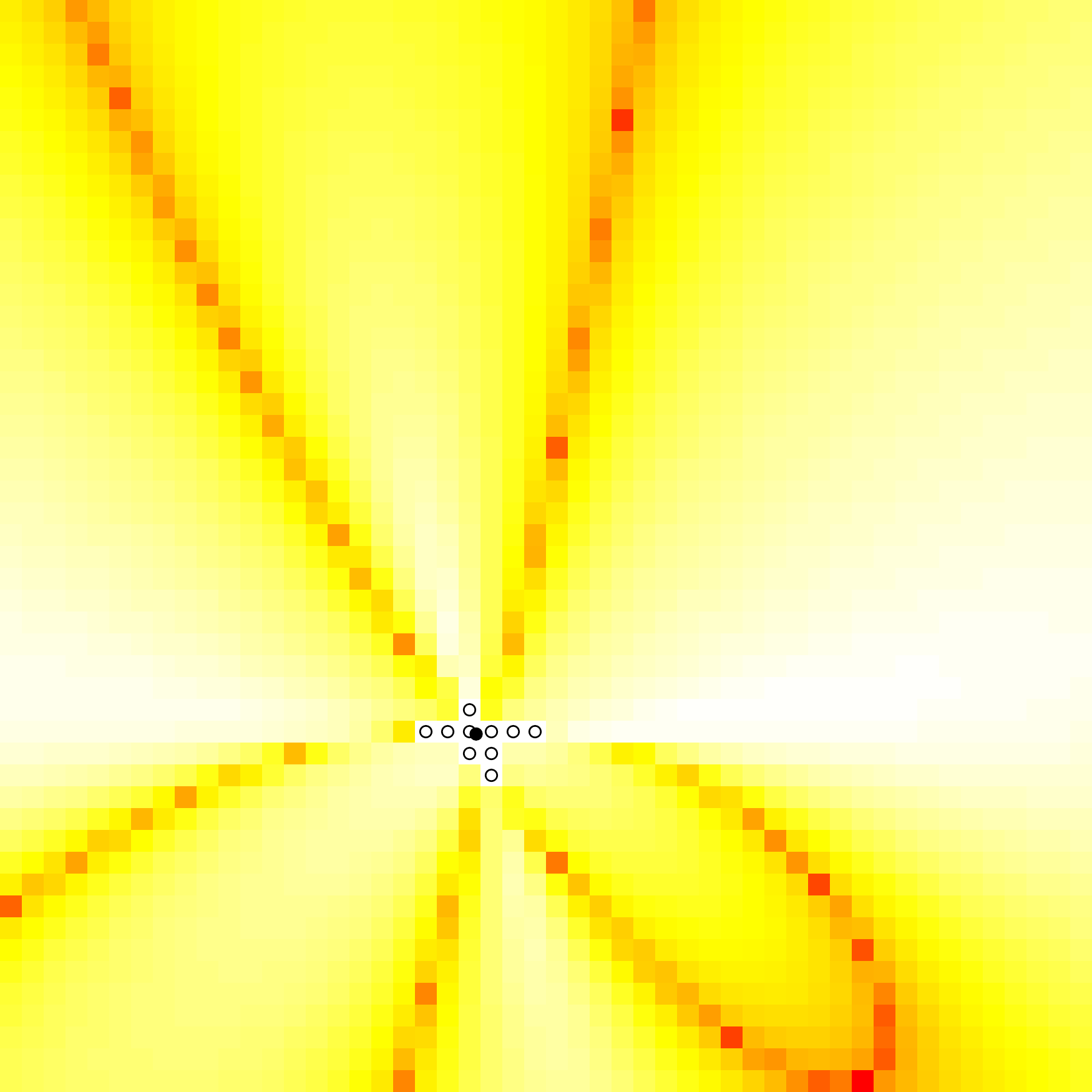} & 
\includegraphics[scale=0.225,trim=15 0 15 0]{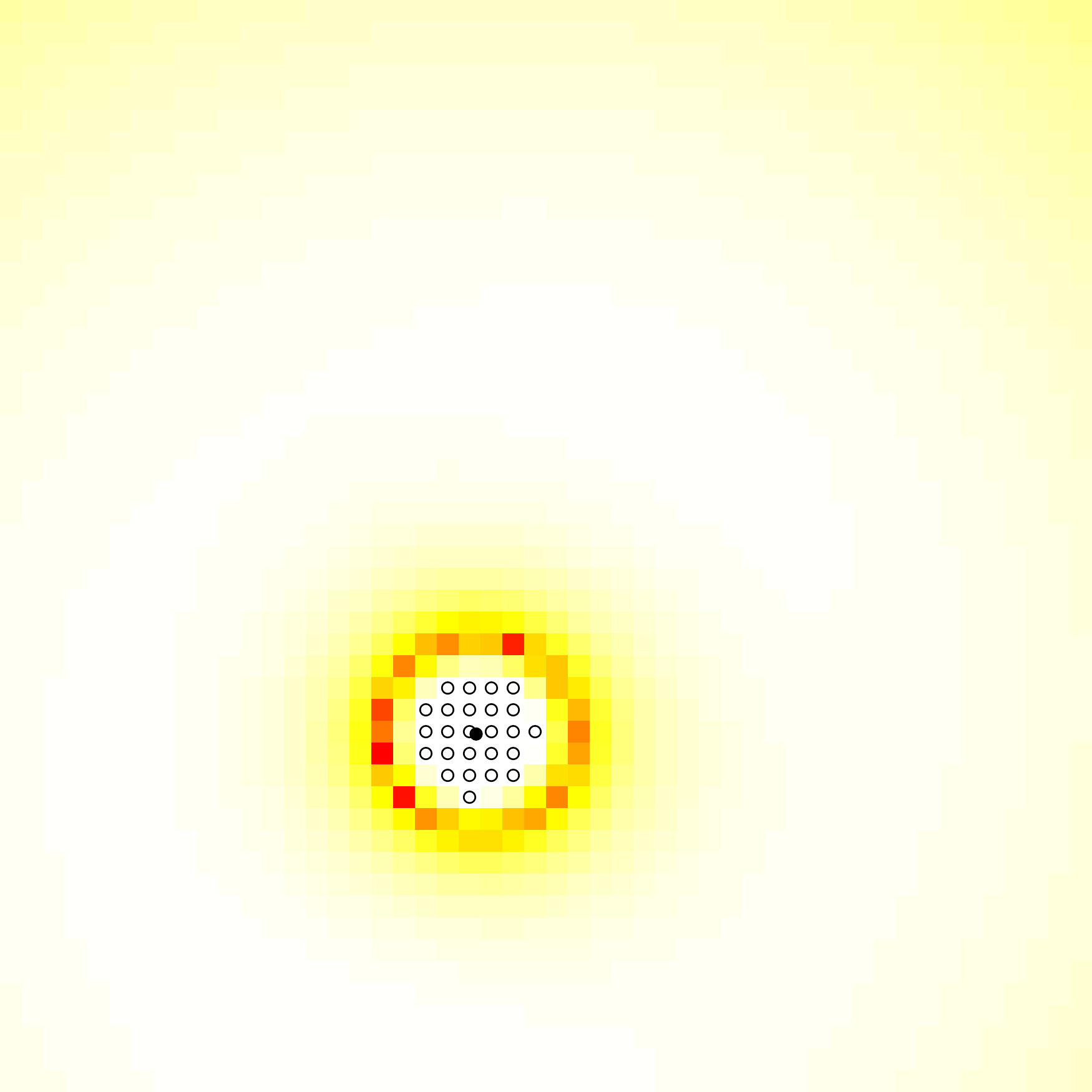} &
\includegraphics[scale=0.225,trim=15 0 15 0]{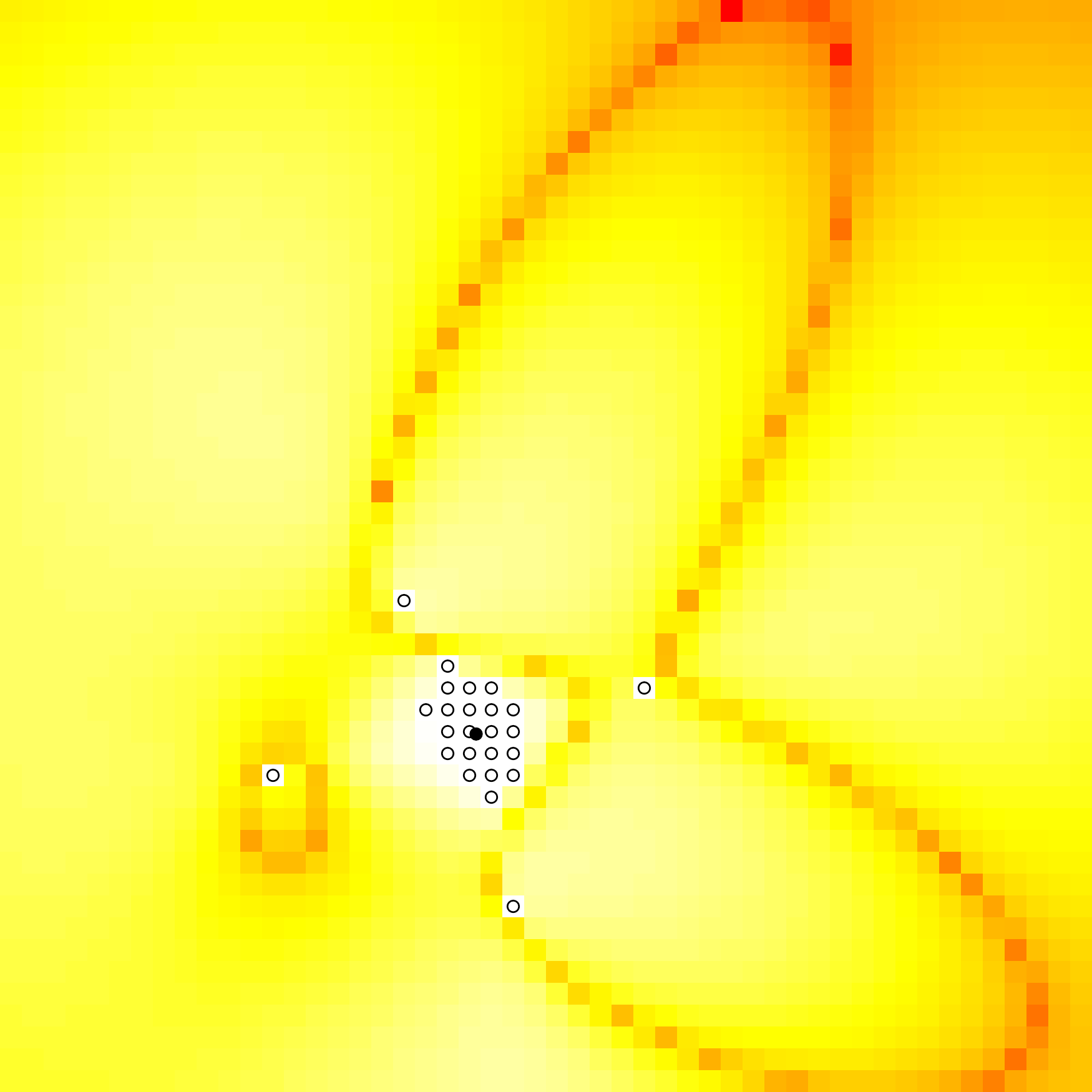} &
\includegraphics[scale=0.225,trim=15 0 0 0]{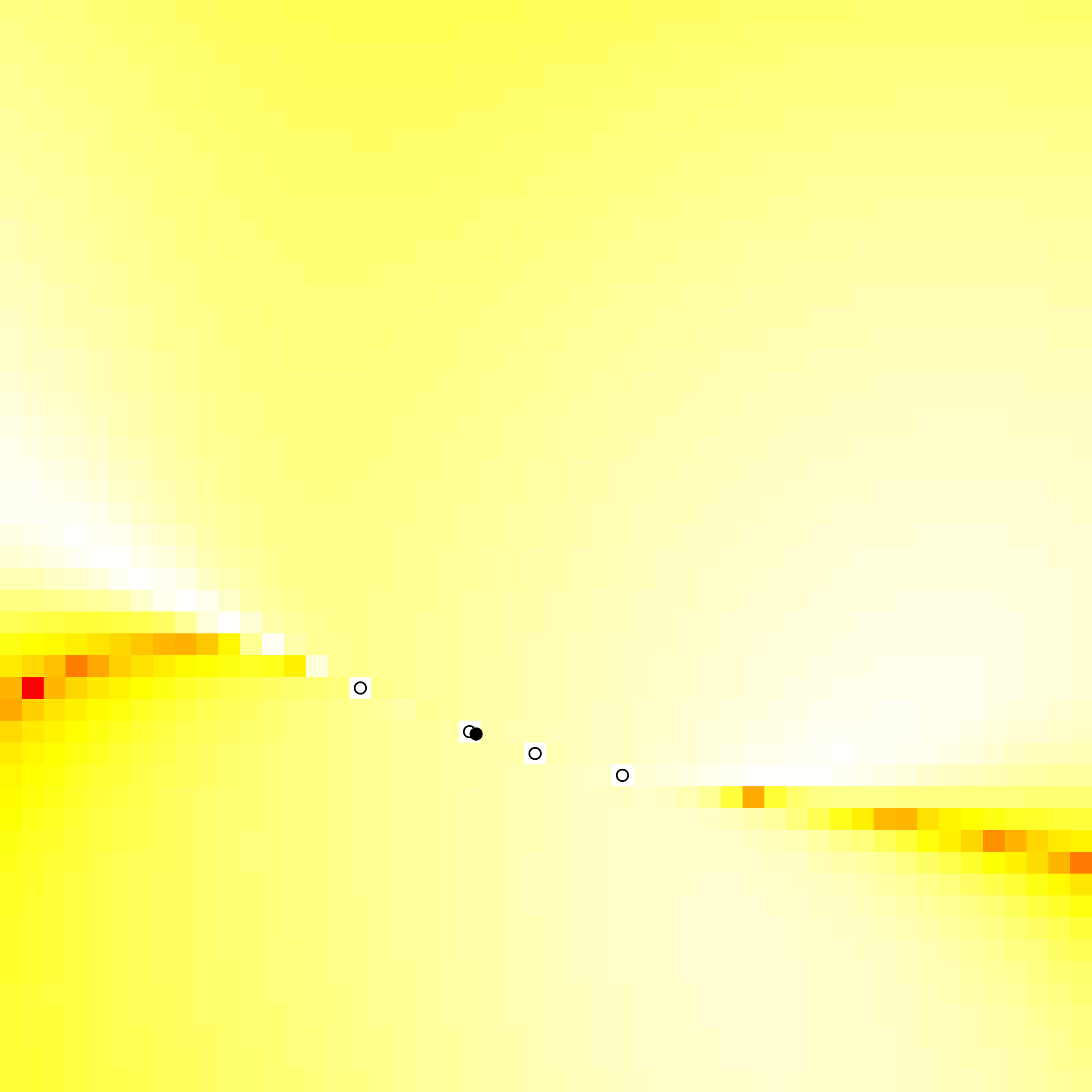}
\\
\includegraphics[scale=0.225,trim=35 0 15 0]{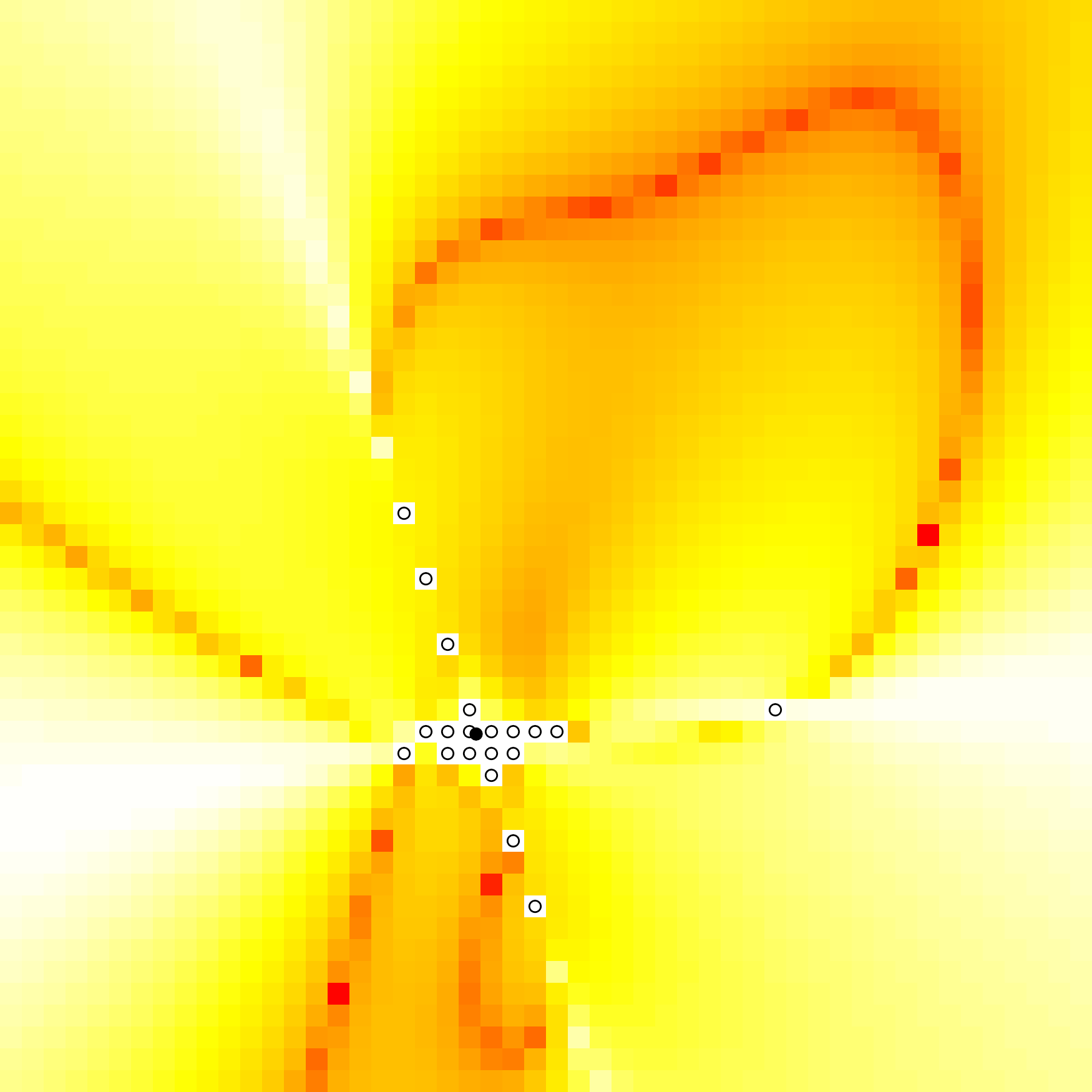} & 
\includegraphics[scale=0.225,trim=15 0 15 0]{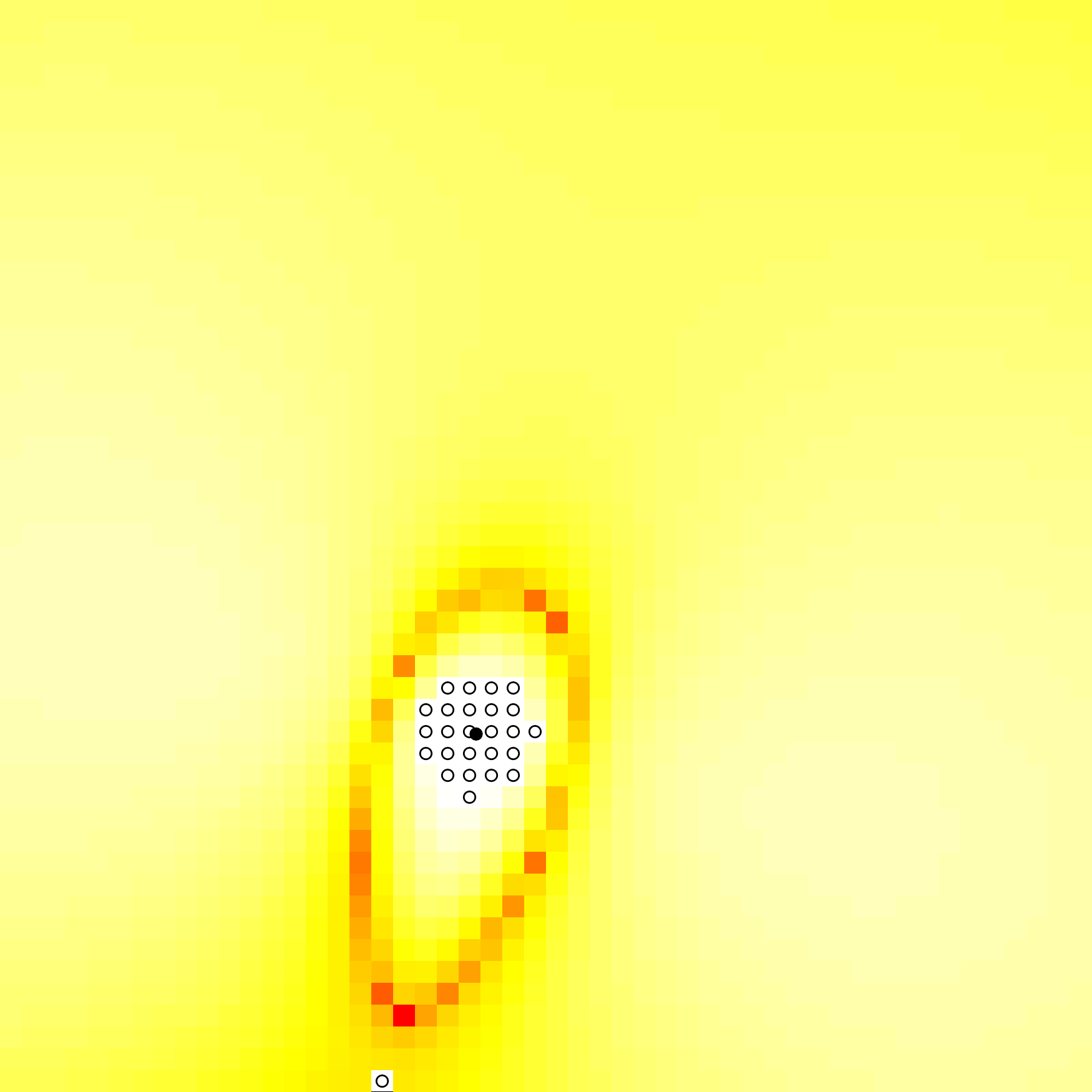} &
\includegraphics[scale=0.225,trim=15 0 15 0]{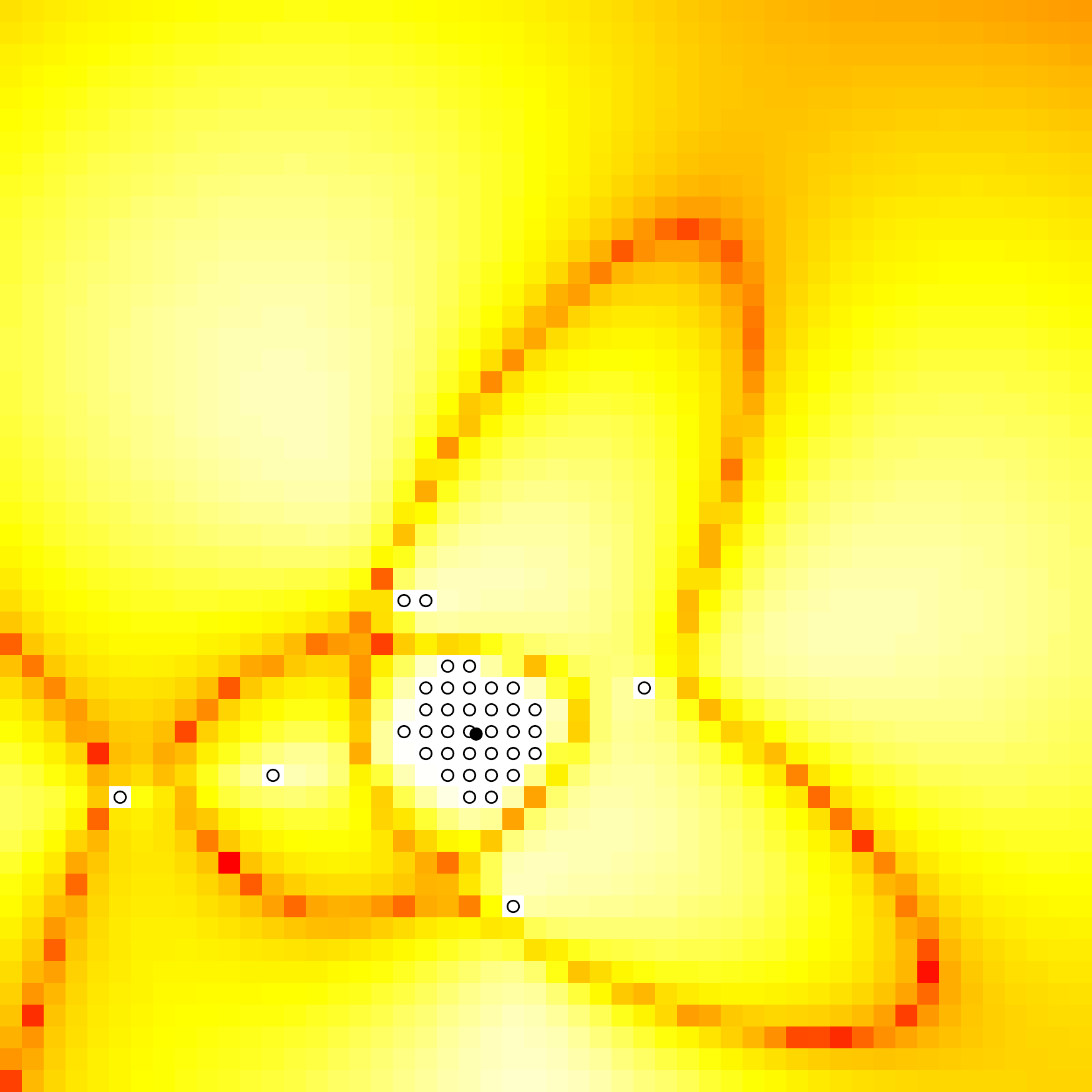} &
\includegraphics[scale=0.225,trim=15 0 0 0]{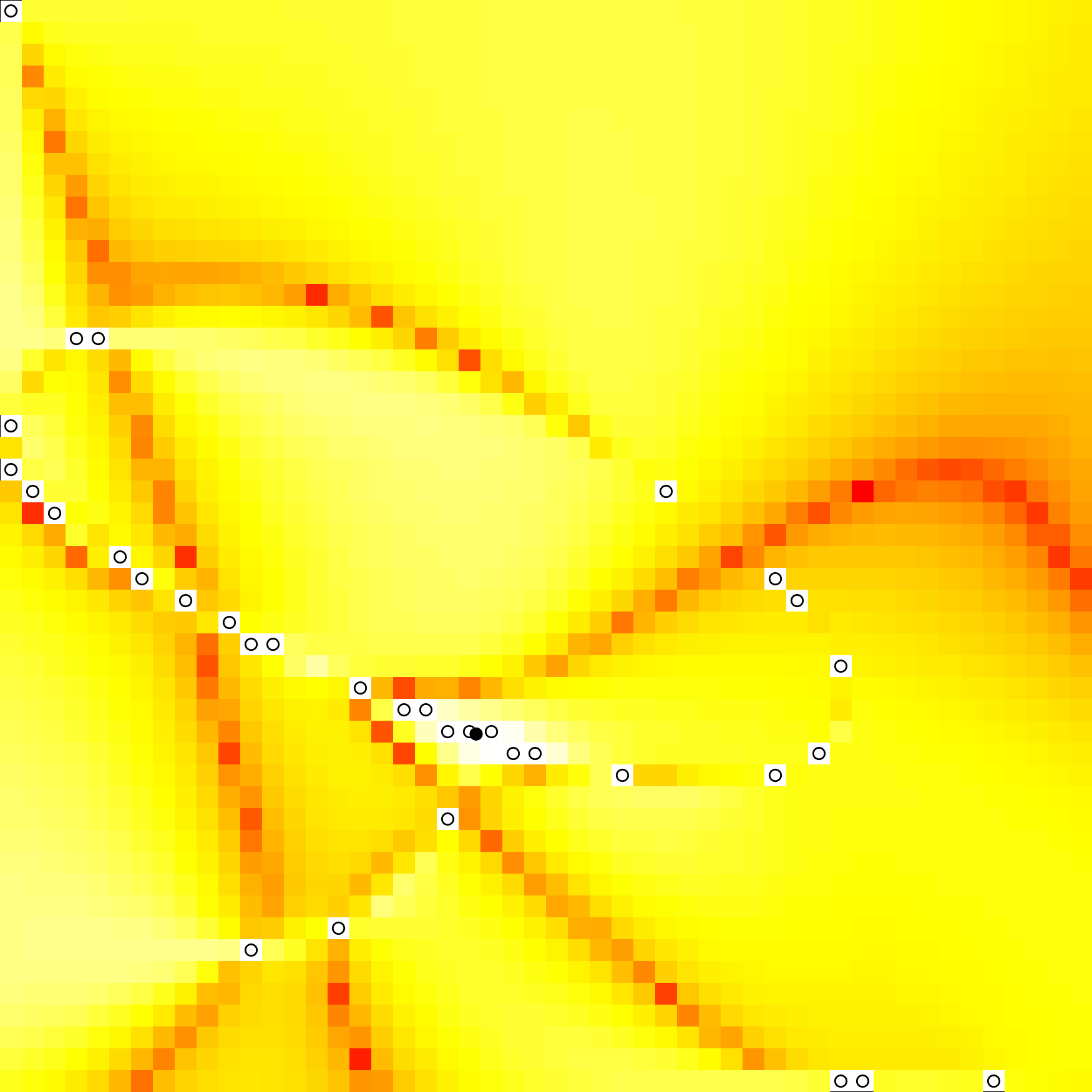} 
\\
\includegraphics[scale=0.225,trim=35 10 15 0]{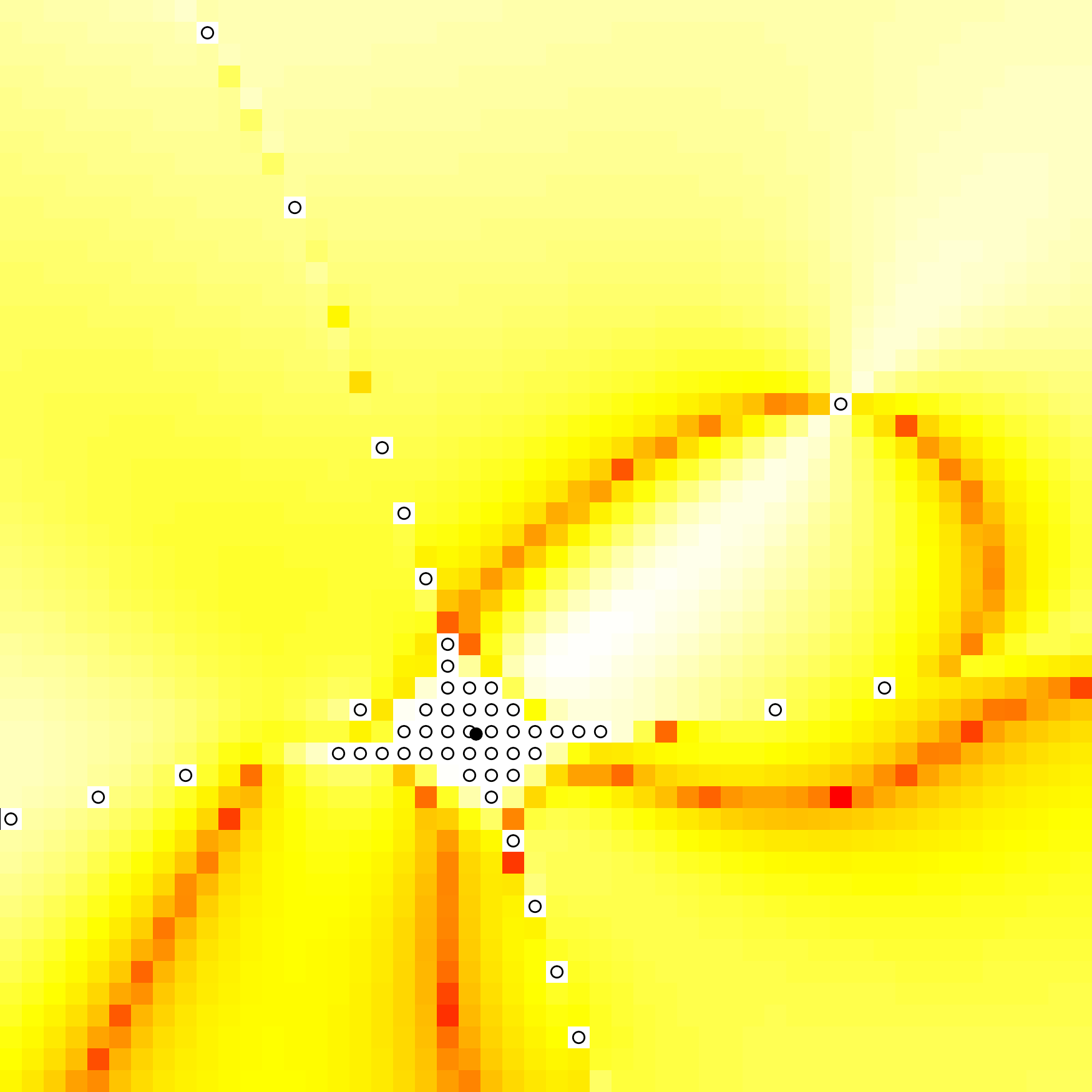} & 
\includegraphics[scale=0.225,trim=15 10 15 0]{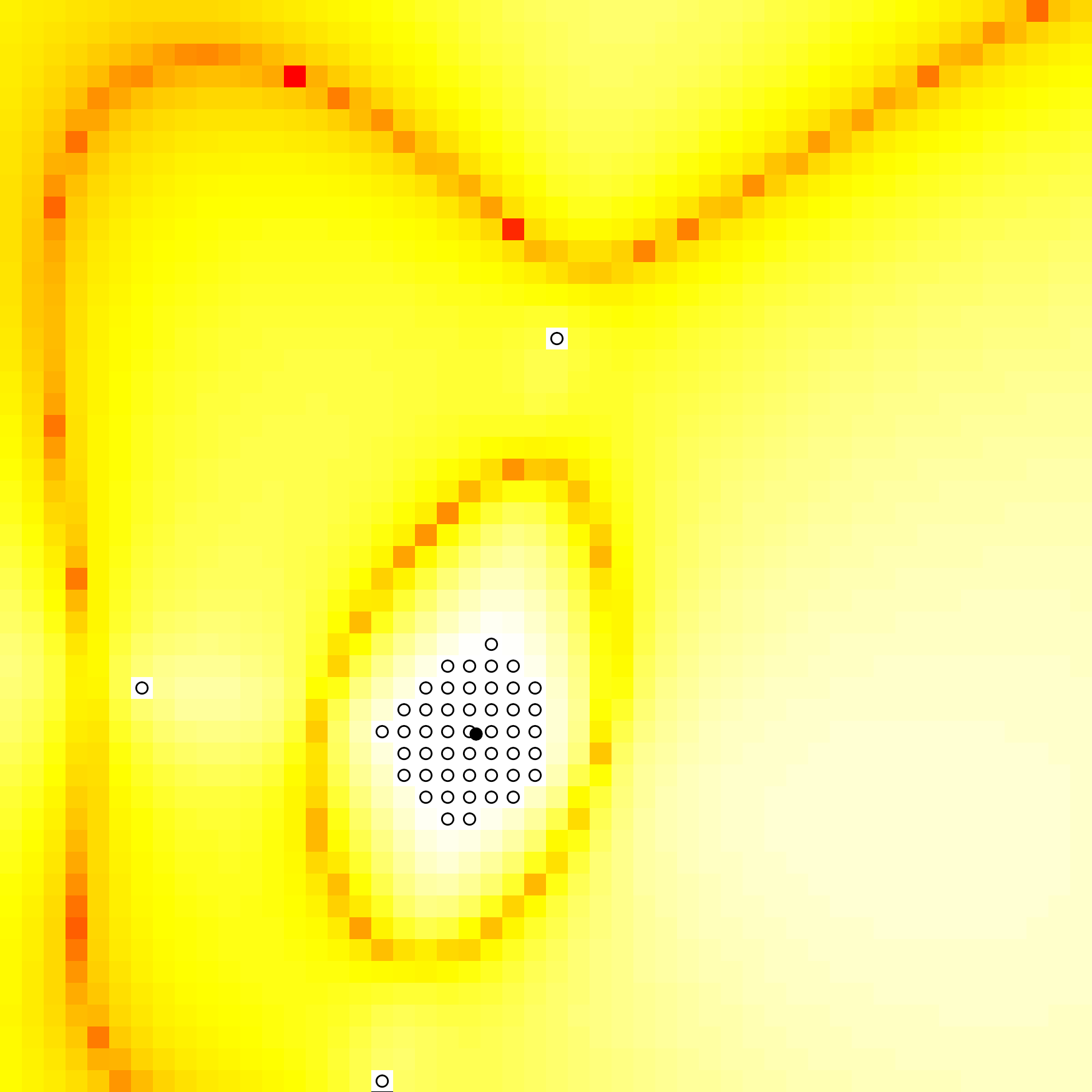} &
\includegraphics[scale=0.225,trim=15 10 15 0]{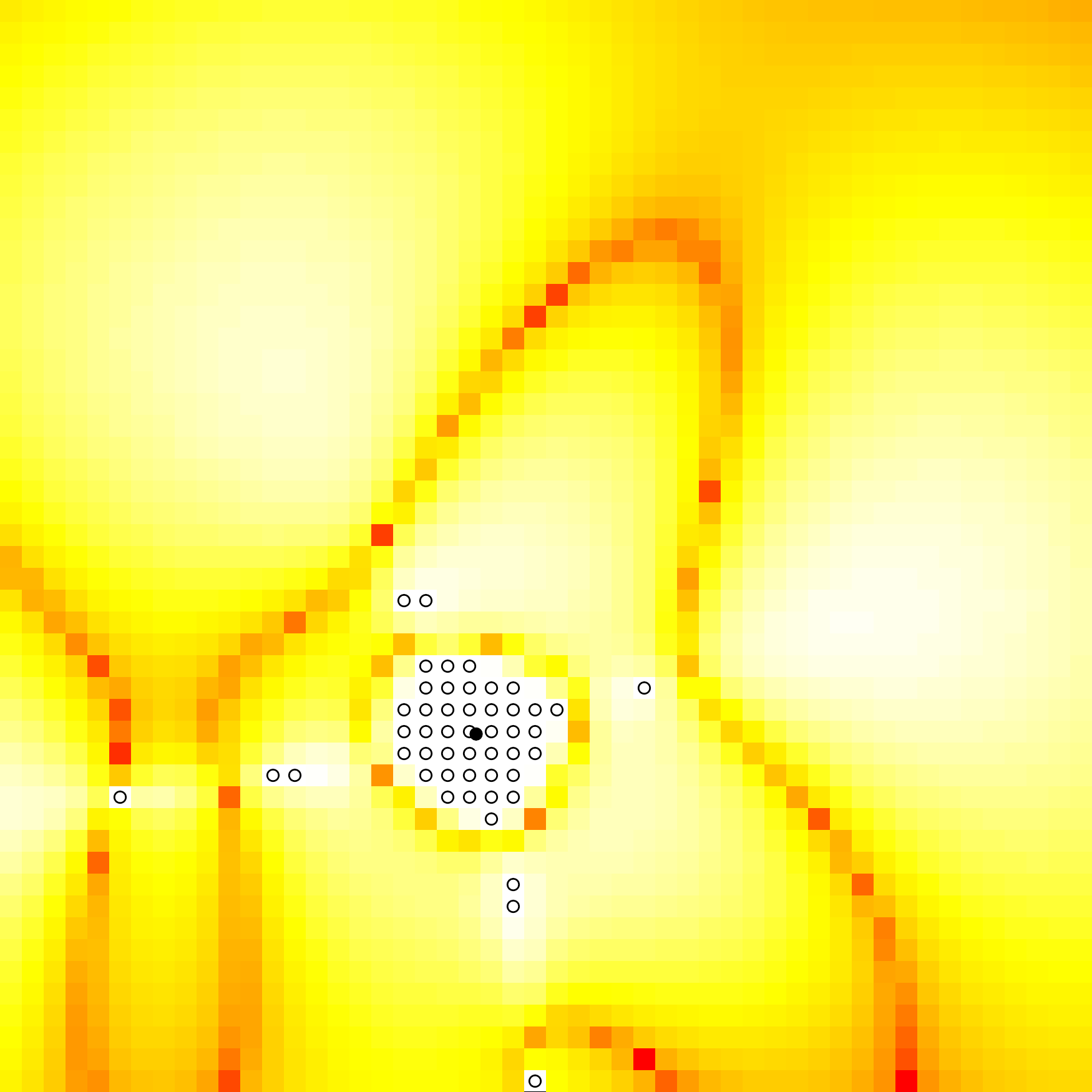} & 
\includegraphics[scale=0.225,trim=15 10 0 0]{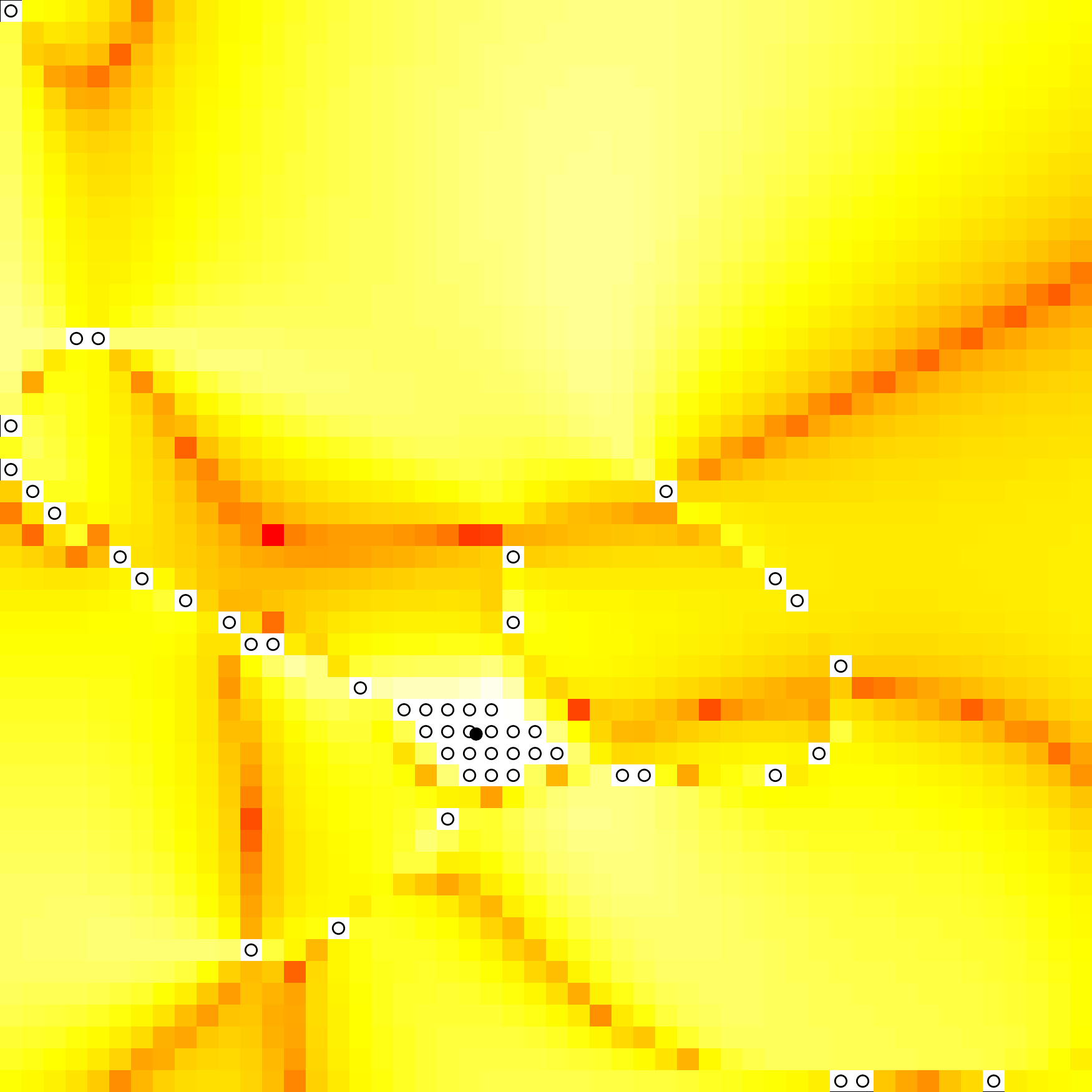}
\end{tabular}
\caption{Progress and reduction in variance surfaces at various stages of the search under
different initial conditions and GP parameters.  Whites are higher values;
reds lower.  The four snapshots in each column were taken at
$j=\{9,10,20,50\}$; $j=\{9,25,26,50\}$; $j=\{24,25,38,49\}$; and
$j=\{1,4,30,50\}$, respectively.  The full sequence for each case is provided
as supplementary material.}
\label{f:snap}
\end{figure}
Figure \ref{f:snap} shows snapshots along the trajectory of four separate
greedy searches (one in each column) for prediction at the same $x$
location, shown as a solid dot.  The image plots show the reduction in
variance criteria [integrand of (\ref{eq:alc})], with lighter shades being
higher values, and open circles representing the current local design
$X_j(x)$.  The full set of plots for each of $n=50$ greedy selections are
provided as supplementary material.

First some high level observations.  The red ribbons and rings, carving out
troughs of undesirable locations for future sampling, are fascinating. They
persist, with some refinement, as greedy selections progress (down the rows),
choosing elements of $X_j(x)$, the open circles. Moreover these selections,
whether near to $x$ or far, impact future choices spanning great distances in
both space and time, i.e., impacting $x_{j+k}$ many iterations $k$ into the
future.

Now some lower level observations along the columns.  Focusing on the first
one, notice how the local design initially spans out in four rays emanating
from $x$.  But eventually points off those rays are chosen, possibly forming
new rays.  The third column shows a similar progression, although the larger
$\eta$ combined with smaller $\theta$ leads to a bigger proportion of NNs.
The first and second rows in those columns show successive designs for $j$ and
$j+1$ (however with different $j$), and it is interesting to see how those
choices impact the resulting reduction in variance surface.  In the first
column, the choices ``tie'' a ribbon of red closer to $x$ in one corner of the
design space.  In the third, where the same behavior might be expected,
instead the surface flattens out and a red ring is created around the new
point.

The second and fourth column show more extreme behavior.  In the second the
large $\theta$ and $\eta$ lead to many NNs being chosen.  However, eventually
the design does fan out along rays, mimicking the behavior of the first and
third columns. The fourth column stands out as exceptional.  That search is
initialized at $n_0 = 1$ and uses a nearly-zero nugget $\eta$.   The first
dozen or so locations follow an arc emanating from $x$, shown in the
first two rows.  Eventually, in the third row, the arc circles around
on itself, creating a ring of points.  However now some points off of the ring
have also been selected, both near and far.  The final row shows the ring
converting into a spiral.  More importantly, it shows a local design starting
to resemble those from the other columns: NNs with satellite points positioned
(loosely) along rays.

Our takeaway from this empirical analysis is that the pattern of local designs
is robust to the choice of correlation parameters, although the proportion of
NNs to satellite points, and the number of rays and their orientation is
sensitive to those choices.  We worked with a gridded design,
which lent a degree of regularity to the illustration.  In other experiments
(not shown) we saw similar behavior as long as $X_N$ was space-filling.
However, obviously if $X_N$ does not accommodate selecting local designs along
rays, then an $X_n(x)$ could not exhibit them.  Below we 
suggest a searching strategy leveraging the patterns observed here, but
which does not preclude discovering different local structure in a less idealized
setting.

\section{Exploiting local influence for efficient search}
\label{sec:exploit}

Here we exploit the partition between NNs and satellite points placed along
rays observed in Section \ref{sec:rr}.  We replace an exhaustive search
amongst candidates $X_N \setminus X_j(x)$ with a continuous line search that
can be solved quickly via a standard 1-d numerical optimizer, and then
``snap'' the solution back to the nearest element of the candidate set.
Recognizing that the greedy local design pattern is sensitive to
initialization and parameterization, as exemplified in Figure \ref{f:snap},
our scheme is conservative about choosing search directions: looking along
rays but not emphasizing the extension of existing ones. 


Consider an existing local design $X_j(x)$ and a search for $x_{j+1}$ along rays 
emanating from $x$.  
\begin{figure}[ht!]
\centering
\includegraphics[scale=0.6,trim=10 20 0 50]{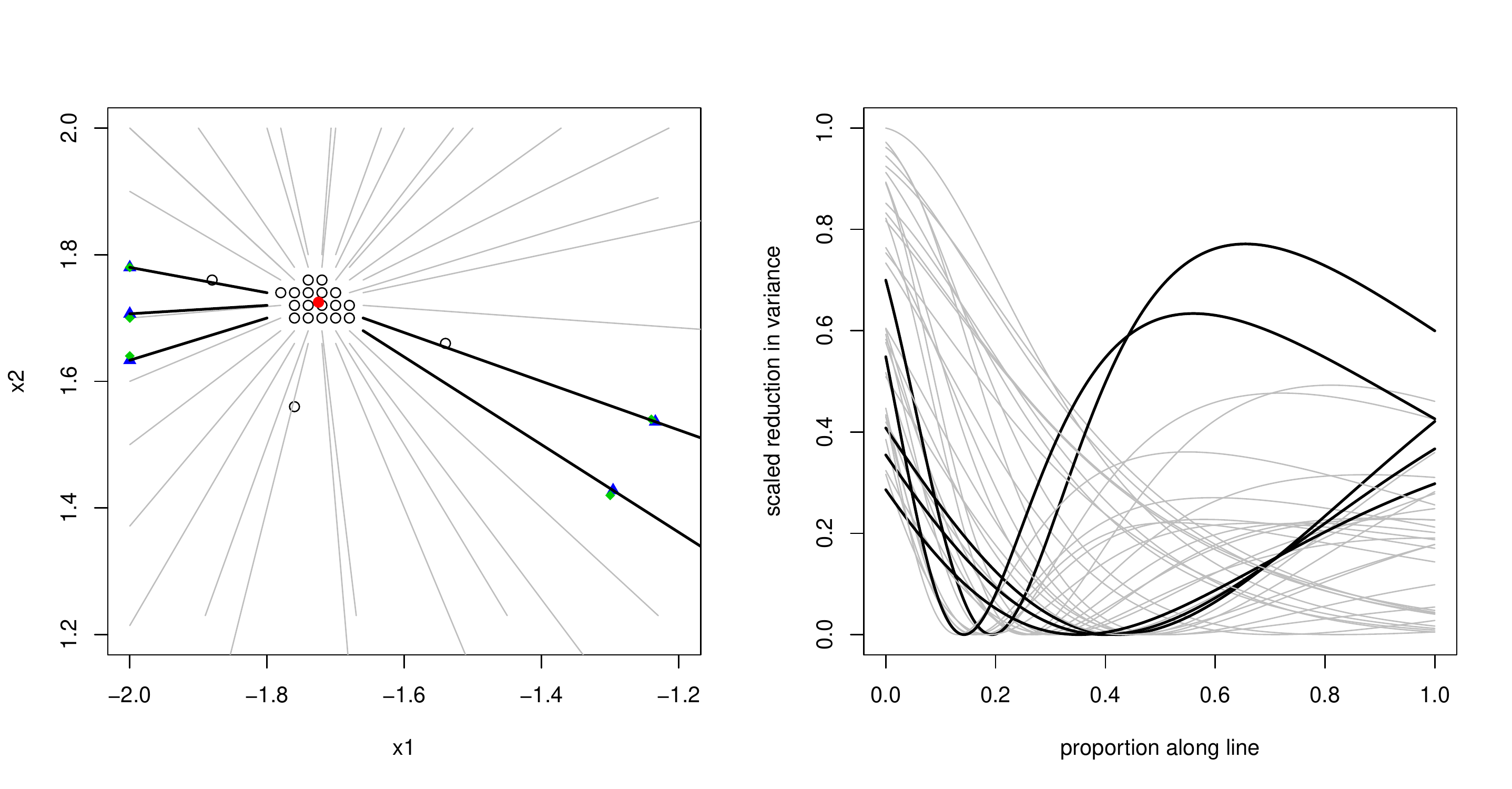}
\caption{Searching for $x_{j+1}$ along rays emanating from $x$.  The {\em
left} panel shows the rays; the {\em right} panel shows the reduction in
variance statistic along those rays.  Gray rays/lines indicate that the
reduction in variance is maximized at its origin, nearest to $x$.  Black rays
indicate maximization farther away, at the location indicated by the blue
triangles.  The red dot is $x$ and the open circles are $X_{20}(x)$.  The
green diamonds are the nearest element $X_N \setminus X_{20}(x)$ to the blue
triangle.}
\label{f:rays}
\end{figure}
For example, the {\em left} panel of Figure \ref{f:rays} shows rays at
iteration $j=20$.  Determining the start and end of the rays---i.e., the
actual line segments---is discussed shortly.  For each ray the corresponding
1-d graph of reduction in variance along the segment is shown in the {\em
right} panel.  Of the thirty five rays, five have a reduction in variance
which is maximized away from its origin, at $x$, and therefore away from the
NN set. Those are shown in black, with blue triangles denoting the optima, and
green diamonds indicating the closest element of the candidate set, $X_N
\setminus X_{20}(x)$, to that point.  The {\em right} panel shows that if a
random ray were chosen, the chances would be $5/35$ that a non NN point would
be selected. For reference we note that the next satellite point is
chosen by exhaustive search at $j=27$.

We select the nearest candidates in $X_N \setminus X_j(x)$ to $x$ as potential
starting points for a ray shooting in the opposite direction.  There is no
need to search between those nearest candidates and $x$ since there are, by
definition, no remaining candidates there.  We allow the segment to stretch to
ten times the distance between the starting point and $x$, or to the edge of
the bounding box of the full candidate set $X_N$.  The rays in Figure
\ref{f:rays} follow this rule. This choice allows the segments to grow in
length as NNs accumulate in $X_j(x)$; both the starts and ends of rays will be
spaced farther from each other and from $x$ as $j$ increases. Although we show
many rays in the figure, only a select few are considered in each iteration
$j$.  These are based on a pseudo-random sequence on the NN candidates,
following
\[
k = (j-n_0+1)\!\!\!\mod \lfloor \sqrt{j-n_0+1} \rfloor.
\]
Then, if the dimension of the input space is $p$, we take $k^\mathrm{th},
k+1^\mathrm{st}, \dots, k+p^\mathrm{th}$ closest elements to $x$ in $X_N
\setminus X_n(x)$ as starting points for ray searches. The start of the
sequence, $k$, is designed to be a round robin over the nearest elements, where
the scope of the ``round'' increases as $j$ increases, discouraging
successive searches from identical starting points. Creating
$p$ rays acknowledges that as the input dimension grows so too does
the complexity of search, however number of rays could be a user-defined
choice adjusting the speed (fewer rays is faster) and fidelity (but also
cruder) as the situation recommends. We note that ray starting points could
also be chosen randomly, however that would result in a stochastic local
design $X_n(x)$, and therefore stochastic predictions, which would be
undesirable in many contexts.

Any 1-d optimization scheme can be used to search over the convex combination
parameter $s \in [0,1]$, spanning the endpoints of the line segment.  We use
the derivative-free \verb!Brent_fmin! method
\citep{brent:1973}, which is the workhorse behind {\sf R}'s {\tt optimize}
function, leveraging a {\sf C} implementation of {\sf Fortran} library
routines (\url{http://www.netlib.org/fmm/fmin.f}).  A bespoke derivative-based
method could speed convergence, however we prefer the well-engineered library
routine in our setting of repeated application over $j=n_0, \dots,
n$ for each of a potentially huge set of $x$'s. We recognize that
\verb!Brent_fmin!'s' initialization biases the solver towards solutions in
the interior of the search space, while we observe [see, e.g., Figures
\ref{f:snap}--\ref{f:rays}] that there is nearly always a mode at
the NN ($s = 0$), in addition to possibly multiple others.  Therefore we check
against the $s=0$ mode as a post-processing step.

Optimizing over rays replaces an exhaustive discrete search with a continuous one,
providing a solution $x_{j+1}^*$ off the candidate set.  So the final step
involves ``snapping'' $x_{j+1}^*$ onto $x_{j+1} \in X_N \setminus X_j(x)$, by
minimizing Euclidean distance.  When performing that search, we explicitly
forbid the ray starting location(s) from being chosen, unless its location
agrees precisely with $x_{j+1}^*$, i.e., unless \verb!Brent_fmin! returns
$s=0$. This prevents choosing a NN location known {\em not} to maximize the
search objective criteria.

Besides these modifications---pseudo random round-robin ray choice, numerical
optimization over a convex combination, and snapping back to the candidate
set---the proposed scheme is identical to the one described in Section
\ref{sec:lagp}.  The green choices shown in Figure \ref{f:alc50} show that the
local designs based on ray search are qualitatively similar to the exhaustive
ones.  What remains is to demonstrate comparable out of sample predictive
performance at substantially reduced computational cost.

\section{Implementation and empirical comparison}
\label{sec:illus}

The methodological developments described above, and all local GP comparators,
are implemented in the {\tt laGP} package on CRAN. The code is primarily in
{\sf C} and utilizes {\tt OpenMP} for multi-core parallelization, as
described by \citet{gramacy:apley:2014}.  Our experiments distribute the local
design over eight threads on a four-core hyperthreaded 2010-model iMac.
GPU subroutines for parallelized exhaustive search
of the reduction in variance criteria are also provided in the package,
however we do not include any new runs leveraging that feature in the empirical
work reported here.  We will, however, compare to the timings on GPUs and
multi-node implementations described by
\citet{gramacy:niemi:weiss:2014}.

All local designs are coupled with local inference for the lengthscale,
$\hat{\theta}_n(x)|X_n(x)$, via Newton-like maximization of the local
posterior probabilities. We also consider second stage re-designs where
sub-designs were re-calculated based on the local parameter estimates obtained
from the first sub-designs, i.e.,  computing
$\tilde{X}_n(x)|\hat{\theta}_n(x)$ after $\hat{\theta}_n(x)|\tilde{X}_n(x)$.
NN comparators do not benefit from a second-stage re-design, since the NN set
is independent of $\theta$.  All exhaustive search variations use a limited
set of NN candidates for each $x$ of size $N' \ll N$ in order to keep
computational costs manageable for those comparators. We primarily use $N' =
1000$, however there are some exceptions to match experiments reported in
earlier work.  No such limits are placed on ray based searches, however
snapping to the nearest element of the candidate set could be sped up slightly
with similarly narrowed search scope.  Unless an exception is noted, each
ray-based search in our experiments uses $p$ rays for $p$-dimensional $X_N$,
however in the software this is a knob that can be adjusted by the user.


\subsection{A synthetic 2-d data set}

Consider a synthetic 2-d dataset, whose $201\times 201$ gridded input design
$X_{N=40401}$ in $[-2,2]^2$ was used for illustrations earlier in the paper.
The response follows $f(x_1,x_2) = -w(x_1)w(x_2)$, where $w(x) =
\exp\left(-(x-1)^2\right) + \exp\left(-0.8(x+1)^2\right)
- 0.05\sin\left(8(x+0.1)\right)$.  
\begin{table}[ht!]
\centering
\begin{minipage}{9cm}
\begin{tabular}{lrr|rrrr}
\multicolumn{6}{c}{iMac} \\
  \hline
meth & $n$ & stage & secs & RMSE & SD & 95\%c \\ 
  \hline
  ray & 200 & 2 & 2948.9 & 0.0002 & 0.0048 & 1.00 \\ 
  ray & 200 & 1 & 1695.4 & 0.0004 & 0.0047 & 1.00 \\ 
  ray & 50 & 2 & 87.8 & 0.0008 & 0.0061 & 1.00 \\ 
  ext & 50 & 2 & 436.8 & 0.0008 & 0.0058 & 1.00 \\ 
  ray & 50 & 1 & 45.6 & 0.0009 & 0.0061 & 1.00 \\ 
  nn & 200 &  & 710.3 & 0.0010 & 0.0027 & 1.00 \\ 
  ext & 50 & 1 & 390.4 & 0.0010 & 0.0060 & 1.00 \\ 
  nn & 50 & & 11.5 & 0.0023 & 0.0045 & 1.00 \\ 
  sub & 1000 & & 50.7 & 0.0369 & 0.0398 & 0.94 \\
   \hline
\end{tabular}
\end{minipage}
\hfill
\begin{minipage}{5.6cm}
\begin{tabular}{rr|rr}
& & \multicolumn{2}{c}{seconds} \\
\hline
&& ext & rays \\
$n$ & $N'$ & C/GPU & iMac \\
\hline
50 & 1000 & 91 & 46 \\
50 & 2000 & 120 & 46 \\
128 & 2000 & 590 & 377 \\
\hline
\end{tabular}
\end{minipage}
\caption{Performance of local GP variations on simple 2-d experiment. The rows
of the table on the left are sorted by the (out-of-sample) RMSE column.  As a
point of reference, a mean-only model has an in-sample RMSE of about 0.2085.
The times reported for 2-stage methods include those from the first stage.  The
table on the right shows the time required for variations on $n$ and $N'$.}
\label{t:2d}
\end{table}
Table \ref{t:2d}, {\em left}, summarizes a predictive experiment involving a
$99\times 99$ grid of $9801$ locations spaced to avoid $X_N$.  Since $X_N$ is
extremely dense in the 2-d space, we opted for just one ray (rather than
$p=2$) at each local design search.  Using two rays gives slight improvements
on RMSE, but requires nearly twice the computation time for the larger $n=200$
experiments. A more in-depth study of how the number of rays is related to
accuracy is deferred to Section \ref{sec:more}.  The most important result
from the table is that the ray-based method requires about 10--20\% of the time
compared to the exhaustive (``ext'') search, and performs at least as well by
RMSE. Design by rays is also more accurate, and 10x faster than, a large NN
comparator.  The simple option of randomly sub-sampling a design of size 1000
(``sub''), and then performing inference and prediction under an isotropic
model, requires commensurate computing resources but leads to poor accuracy
results.\footnote{Estimating a separable global model on a subset of the data
leads to similar accuracy/coverage results, since the data-generating
mechanism has a high degree of radial symmetry, at greater computational
expense required to numerically optimize over a vectorized $\theta$
parameter.} Choosing the subset randomly also adds variability: a 95\%
interval for RMSE over thirty repeats is (0.0007, 0.0416).  Space-filling
sub-designs can substantially reduce that variability but at potentially great
computational cost.  For example, calculating an 1000-sized maximum entropy
sub-design would cause compute times to go up by at least two orders of
magnitude (with no change in mean performance)  due to the multitude of
$K_{1000}^{-1}$ calculations that would require.

Some other observations from the {\em left}-hand table include that a second
stage design consistently improves accuracy, and that a large ($n=200$)
search, which would require unreasonable computation (without a GPU) if
exhaustive, is feasible via rays and leads to the best overall predictors.
Although some methods report lower predictive SD than others, note that local
approximations yield perfect coverage, which is typical with deterministic
computer experiment data.  The ``sub'' comparator achieves coverage close to
the nominal 95\% rate, but that ought not impress considering its poor RMSE
results.  Achieving nominal coverage is easy at the expense of accuracy,
e.g., via constant $\mu$ and $\sigma^2$ under a global Gaussian model.
Achieving a high degree of accuracy at the expense of over-covering, and hence
obtaining a conservative estimate of uncertainty, is a sensible tradeoff in
our context.

\citet{gramacy:niemi:weiss:2014} reported relative speedup times for a variant
on the above experiment utilizing dual-socket 8-core 2.6 GHz 2013-model
Intel Sandy Bridge compute cores, and up to two connected Nvidia Tesla M2090 GPU
devices which massively parallelized the exhaustive search subroutine. The
best times from those experiments, involving sixteen threads and both GPUs and
{\em one}-stage design, are quoted on the {\em right} in Table \ref{t:2d}. The
accuracy of the predictions follow trends from the {\em left} table, however
the timing information reveals that searching via rays is competitive with the
exhaustive search even when that search subroutine is offloaded to a GPU. As
the problem gets bigger ($n=128, N'=2000$), the GPU exhaustive search and CPU
ray-based search are competitive, but the ray method does not require limiting
search to the $N'$ nearby locations.

\subsection{Langley Glide-back booster}

\citeauthor{gramacy:niemi:weiss:2014} presented timing results
for emulation of a computer experiment simulating the lift of a rocket booster
as it re-enters the atmosphere after disengaging from its payload.  The design
involved 37908 3-dimensional input settings and a predictive grid of size
644436.  To demonstrate the supercomputing capability of local GPs on a
problem of this size, they deploy exhaustive local search distributed over
four identical 16-CPU-core/2-GPU nodes via the {\tt parallel} library (in
addition to {\tt OpenMP}/{\tt CUDA}) in {\sf R}.  Timings are presented for
several different fidelities of local approximation. These numbers are shown
in Table \ref{t:lgbb} alongside some new numbers of our own from a ray-based
search on the iMac.  

\begin{table}[ht!]
\centering 
\begin{tabular}{rr|rrr|r}
& & \multicolumn{3}{c}{minutes} \\
\hline
&& \multicolumn{3}{c|}{exhaust} & rays \\
$n$ & $N'$ & 1x(16-CPU) & 4x(16-CPU) & 4x(16-CPU/2-GPU) & iMac \\
\hline
50 & 1000 & 235 & 58 & 21 & 65 \\
50 & 2000 & 458 & 115 & 33 & 65  \\
50 & 10000 & 1895 & 476 & 112 & 65 \\
128 & 2000 & - & 1832 & 190 & 1201 \\
\hline
\end{tabular}
\caption{Timings (minutes) for prediction via local Gaussian process alternatives
pitting exhaustive search via GPUs and cluster computing vs.~a
ray-based search on a 4-core hyperthreaded iMac.  Certain configurations required
too much time to include in the comparison.}
\label{t:lgbb}
\end{table}

Observe that times for ray search on the iMac are in the ballpark of the
exhaustive searches.  Depending on $n$ and $N'$ one style of search may be
faster than another, however the exhaustive search clearly demands
substantially more in the way of compute cycles. This a real-data experiment
involved a large OOS predictive set for which no true-values of the output are
known, so we were unable to explore how accuracy increases with fidelity.

\subsection{The borehole data}
\label{sec:large}

The borehole experiment \citep{worley:1987,morris:mitchell:ylvisaker:1993}
involves an 8-dimensional input space, and our use of it here follows the
setup of \cite{kaufman:etal:2012}; more details can be found therein.  We
perform two similar experiments.  In the first one, summarized in Table
\ref{t:borehole}, we duplicate the experiment of
\citet{gramacy:niemi:weiss:2014} and perform out-of-sample prediction based on
designs of increasing size $N$.  The designs and predictive sets (also of size
$N$) are from a joint random Latin hypercube sample.  As $N$ is increased so
is the local design size $n$ so that there is steady reduction of
out-of-sample MSE down the rows of the table.
\citeauthor{gramacy:niemi:weiss:2014} also increase $N'$, the size of the local
candidate set, with $N$, however the ray searches are not limited in this way.

\begin{table}[ht!]
\centering
\begin{tabular}{rr||rrr|rr||rr|rr}
&& \multicolumn{5}{c||}{exhaustive} & \multicolumn{4}{c}{via rays} \\
\hline
&& \multicolumn{5}{c||}{Intel Sandy Bridge/Nvidia Tesla} & 
\multicolumn{2}{c|}{iMac} & \multicolumn{2}{c}{Intel SB} \\
    &     &      &  \multicolumn{2}{c|}{96x CPU} & \multicolumn{2}{c||}{5x 2 GPUs}
& \multicolumn{2}{c|}{1x(4-core) CPU} & \multicolumn{2}{c}{96x CPU}
    \\
\hline
$N$ & $n$ & $N'$ & seconds & mse & seconds & mse & seconds & mse & seconds & mse\\ 
\hline
1000 & 40 & 100 & 0.48 & 4.88 & 1.95 & 4.63 & 8.00 & 6.30 & 0.39 & 6.38 \\ 
2000 & 42 & 150 & 0.66 & 3.67 & 2.96 & 3.93 & 17.83 & 4.47 & 0.46 & 4.10\\ 
4000 & 44 & 225 & 0.87 & 2.35 & 5.99 & 2.31 & 40.60 & 3.49 & 0.62 & 2.72 \\ 
8000 & 46 & 338 & 1.82 & 1.73 & 13.09 & 1.74 & 96.86 & 2.24 & 1.31 & 1.94 \\ 
16000 & 48 & 507 & 4.01 & 1.25 & 29.48 & 1.28 & 222.41 & 1.58 & 2.30 & 1.38\\ 
32000 & 50 & 760 & 10.02 & 1.01 & 67.08 & 1.00 & 490.94 & 1.14 & 4.65 & 1.01\\ 
64000 & 52 & 1140 & 28.17 & 0.78 & 164.27 & 0.76 & 1076.22 & 0.85 & 9.91 & 0.73\\ 
128000 & 54 & 1710 & 84.00 & 0.60 & 443.70 & 0.60 & 3017.76 & 0.62 & 17.99 & 0.55 \\ 
256000 & 56 & 2565 & 261.90 & 0.46 & 1254.63 & 0.46 & 5430.66 & 0.47 & 40.16 & 0.43 \\ 
512000 & 58 & 3848 & 836.00 & 0.35 & 4015.12 & 0.36 & 12931.86 & 0.35 & 80.93 & 0.33 \\ 
1024000 & 60 & 5772 & 2789.81 & 0.26 & 13694.48 & 0.27 & 32866.95 & 0.27 & 188.88 & 0.26 \\
2048000 & 62 & - & - & - & - & - & - & - &  466.40 & 0.21 \\ 
4096000 & 64 & - & - & - & - & - & - & - &  1215.31 & 0.19 \\ 
8192000 & 66 & - & - & - & - & - & - & - &  4397.26 & 0.17 \\
\hline
\end{tabular}
\caption{Timings and out-of-sample accuracy measures for increasing problem
sizes on the borehole data. The ``mse'' columns are mean-squared predictive
error to the true outputs on the $|\mathcal{X}| = N$ locations from separate
runs (hence the small discrepancies between the two columns).  \ 95\%
coverages are nominaly within 1\% of 1.00 for all comparators, so they are not
reported in the table.  Both CPU and GPU nodes have 16 CPU cores.  So the
``96x CPU'' shorthand in the table indicates 1536 CPU cores.}
\label{t:borehole}
\end{table}

First focus on the timing and accuracy (MSE) results in the ``iMac via rays''
columns.  The four columns to the left summarize the earlier experiment(s). By
comparing the time column we see that the 4-core hyperthreaded iMac/rays
implementation is about 2.5x times slower than the combined effort of 80 cores
and 2 GPUs, or about 11.5x times slower than using more than 1500 cores.  The
amount of computing time for the largest run, at nine hours for more than
one-million predictions on a one-million-sized design, is quite impressive for
a nearly five-year-old desktop, compared to modern supercomputers (in earlier
columns).  However, looking at the accuracy column(s), we see that while the
ray-based search is giving nearly identical results for the largest
emulations, it is less accurate for the smaller ones.  We attribute this to
the size of the input space relative to the density of the smaller design(s),
with two implications: (1) when designs are small, searching exhaustively is
cheap, so a larger candidate set ($N'$) relative to design size ($N$) can be
entertained; (2) at the same time, since the design is sparse in high
dimension when $N$ is small, the rays intersect fewer candidates, reducing the
chances that the $x_{j+1}$ is close to the solution $x_{j+1}^*$ found along
ray(s).  In a large-$p$ small-$N$ setup, it may be best to search
exhaustively.

The final pair of columns in Table \ref{t:borehole} show timings and accuracies
for runs with rays distributed over 96 Intel/Sandy Bridge 16-core
machines.  Observe how the running times are comparable to the exhaustive
version, shown in $4^\mathrm{th}$ column, until the pair $(N=32000,n=50)$.
Neither method is fully utilizing the massively parallel potential
of this supercomputer on ``smaller'' data.  The exhaustive method is more
accurate, if just slightly slower up until this point, so perhaps that method
may be preferred in the ``smaller'' (but still huge for GPs) setting. But then
the timings diverge, with rays showing big efficiency gains and nearly
identical accuracy.  By $(N=1024000,n=60)$ rays emulate more than twenty times
faster.  We then allowed rays to explore larger problems, to see what size
data could be emulated in about an hour, ultimately finding that we can
accommodate an 8x larger experiment while allowing the approximation fidelity
to rise commensurately up to $n=66$.  This computational feat
is unmatched in the computer modeling literature, whether via
GPs or otherwise.

\begin{table}[ht]
\centering
\begin{tabular}{r||rr|rr||r|rr|rr}
& \multicolumn{2}{c|}{subset iso} & \multicolumn{2}{c||}{subset sep} 
& \multicolumn{5}{c}{rays with pre-scaled inputs} \\
\hline 
& \multicolumn{4}{c||}{iMac} && \multicolumn{2}{c|}{iMac} & \multicolumn{2}{c}{Intel SB} \\
& \multicolumn{4}{c||}{1x(4-core) CPU, $n=1000$} && \multicolumn{2}{c|}{1x(4-core) CPU} & 
\multicolumn{2}{c}{96x CPU} \\
\hline
$N$ & seconds & mse & seconds & mse & $n$ & seconds & mse & seconds & mse\\ 
  \hline
1000 & 55.97 & 0.66 & 296.95 & 0.20 & 40 & 8.50 & 0.94 & 0.35 & 0.79 \\ 
2000 & 55.82 & 0.38 & 220.62 & 0.11 & 42 & 19.73 & 0.38 & 0.50 & 0.47 \\ 
4000 & 59.96 & 0.56 & 264.98 & 0.17 & 44 & 45.51 & 0.34 & 0.75 & 0.29 \\ 
8000 & 60.04 & 0.50 & 288.19 & 0.11 & 46 & 111.14 & 0.22 & 1.46 & 0.23 \\
16000 & 70.76 & 0.53 & 225.68 & 0.17 & 48 & 241.08 & 0.20 & 2.63 & 0.19 \\ 
32000 & 89.75 & 0.50 & 362.09 & 0.13 & 50 & 547.10 & 0.16 & 4.86 & 0.15 \\ 
64000 & 130.08 & 0.49 & 383.42 & 0.12 & 52 & 1208.15 & 0.13 & 9.13 & 0.13 \\ 
128000 & 211.07 & 0.45 & 478.19 & 0.12 & 54 & 2732.71 & 0.11 & 19.50 & 0.12 \\
256000 & 375.29 & 0.43 & 660.55 & 0.12 & 56 & 6235.47 & 0.11 & 41.80 & 0.11 \\
512000 & 719.33 & 0.47 & 977.08 & 0.12 & 58 & 15474.62 & 0.10 & 82.56 & 0.10 \\ 
1024000 & 1475.42 & 0.55 & 1690.75 & 0.13 & 60 & 38640.03 & 0.10 & 192.51 & 0.10 \\
2048000 & - & - & - & - & 62 & - & - & 488.15 & 0.11 \\
4096000 & - & - & - & - & 64 & - & - & 1234.56 & 0.11 \\
8192000 & - & - & - & - & 66 & - & - & 4200.07 & 0.11 \\
   \hline
\end{tabular}
\caption{Timings augmenting Table \ref{t:borehole} to include 
global GP emulators built on random $n=1000$ subsets of the full design under
isotropic and separable formulations.  The final columns show how ray-based
local approximations compare with inputs pre-scaled by (the square root of)
the globally estimated lengthscales estimated from the random data subset.}
\label{t:borehole:sub}
\end{table}

We are grateful to a referee for pointing out the importance comparing the
results in Table \ref{t:borehole} to a global GP predictor applied to a subset
of the training data.  In response we created Table
\ref{t:borehole:sub} where columns 2--5 contain timings and MSE calculations
obtained with $n=1000$ random sub-designs under (estimated) isotropic and
separable correlation functions.  Note that although this part of the table is
organized in rows for varying $N$, the results here are (in expectation)
independent of $N$.\footnote{MSE variance decreases slightly down the rows as
the testing set is increasing in size with $N$.} The borehole function is
slowly varying across the input space, and highly stationary, so it is perhaps
not surprising that a global isotropic GP (columns 2--3) fit on a subset
compares favorably to the local ray-based alternative until $N=256,000$. The
borehole function is not uniformly sensitive to its inputs, so a global
isotropic model is a poor fit, see e.g., \citet{gramacy:2014}. The next pair
of columns (4--5) show improved results with a separable model, although
again these are noisy along the rows.

The final pair of columns come from a hybrid separable global model and local
approximate one, via rays. \cite{gramacy:bingham:etal:2014} observed
substantial performance boosts from pre-scaling inputs by the square-root
global lengthscales estimated from data subsets. Although the local nature of
inference and prediction accommodates a limited degree of nonstationarity, and
thus anisotropy, if the data is highly globally anisotropic (as is the case
for the borehole example), pre-scaling offers several benefits.  One is
statistical efficiency: global trends are best estimated on global scales.
Another is computational: rays search radially, and therefore best mimic
exhaustive searches when patterns in the variance reduction surface are
loosely isotropic (as in Figure \ref{f:snap}).  Pre-scaling aligns the two
more closely in that respect; more detailes are provided by
\citet{gramacy:2014}. The final two columns (7--8) show this hybrid approach
outperforming the rest from about $N=128,000$. Progress appears to plateau
past $N \approx 1$~million, due partly to rounding error/statistical noise,
and partly to $n$ growing too slowly relative to $N$ in this setup.

\subsection{Predictive accuracy and computation time}
\label{sec:more}

The comparison above duplicated a simulation setup from earlier work that was
carefully designed to illustrate performance under increasing fidelity of
approximation as data sizes increase.  Here we consider a direct pitting of
accuracy versus compute time in several views.  In addition to the borehole
data we consider the so-called \citet{zhou:1998} simulation and the piston
simulation function
\citep{kenett:zacks:1998,benari:steinberg:2007}.  The Zhou function can be
applied for a range of input dimensions.  Our first experiment, below, follows
\citet{an:owen:2001} in using $p=9$ dimensions. We use the log of the
response as the data-generating process involves products of exponentials that
cause modeling challenges (and numerical instability) for larger $p$. The piston
data is 7-dimensional.  For more details on both benchmarks see the library
compiled by \citet{simulationlib}, an excellent resource for benchmark
problems involving computer simulation and associated applications.

\begin{figure}[ht!]
\centering 
\includegraphics[scale=0.6,trim=0 72 5 0,clip=TRUE]{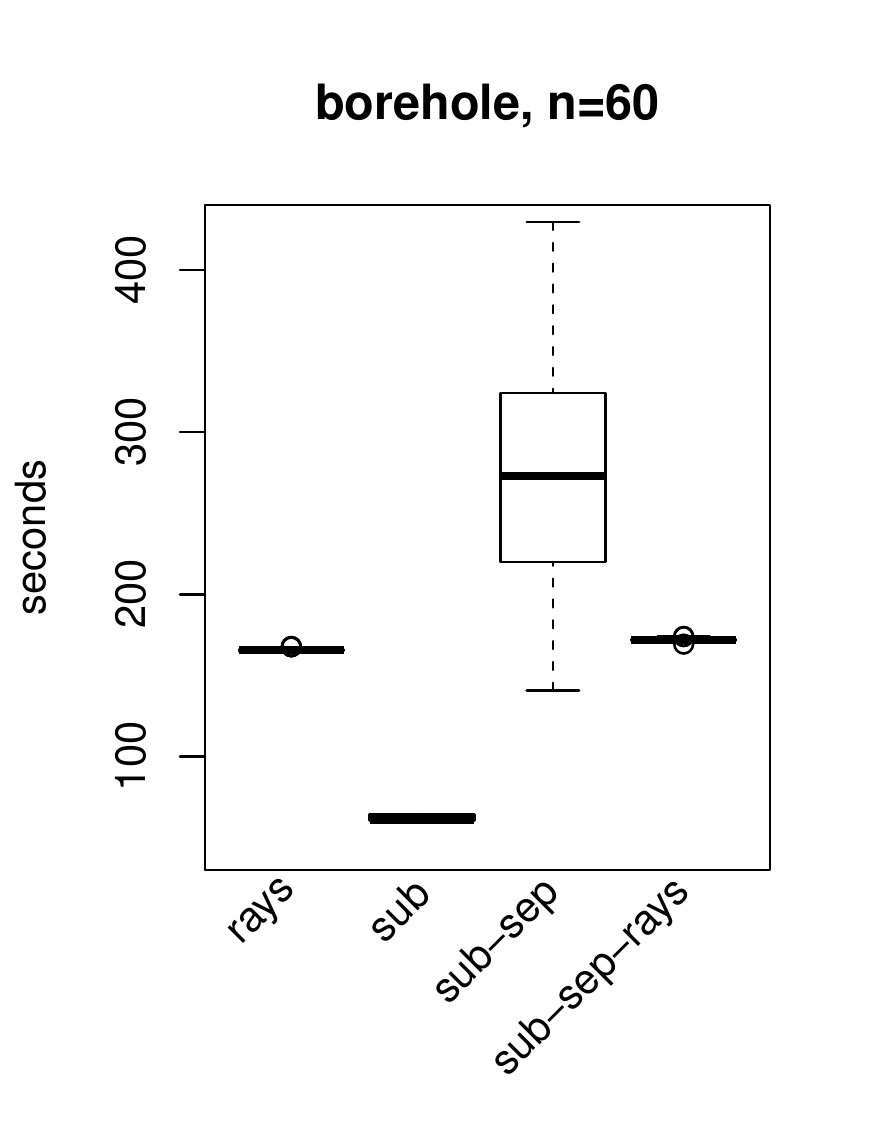}
\includegraphics[scale=0.6,trim=30 72 5 0,clip=TRUE]{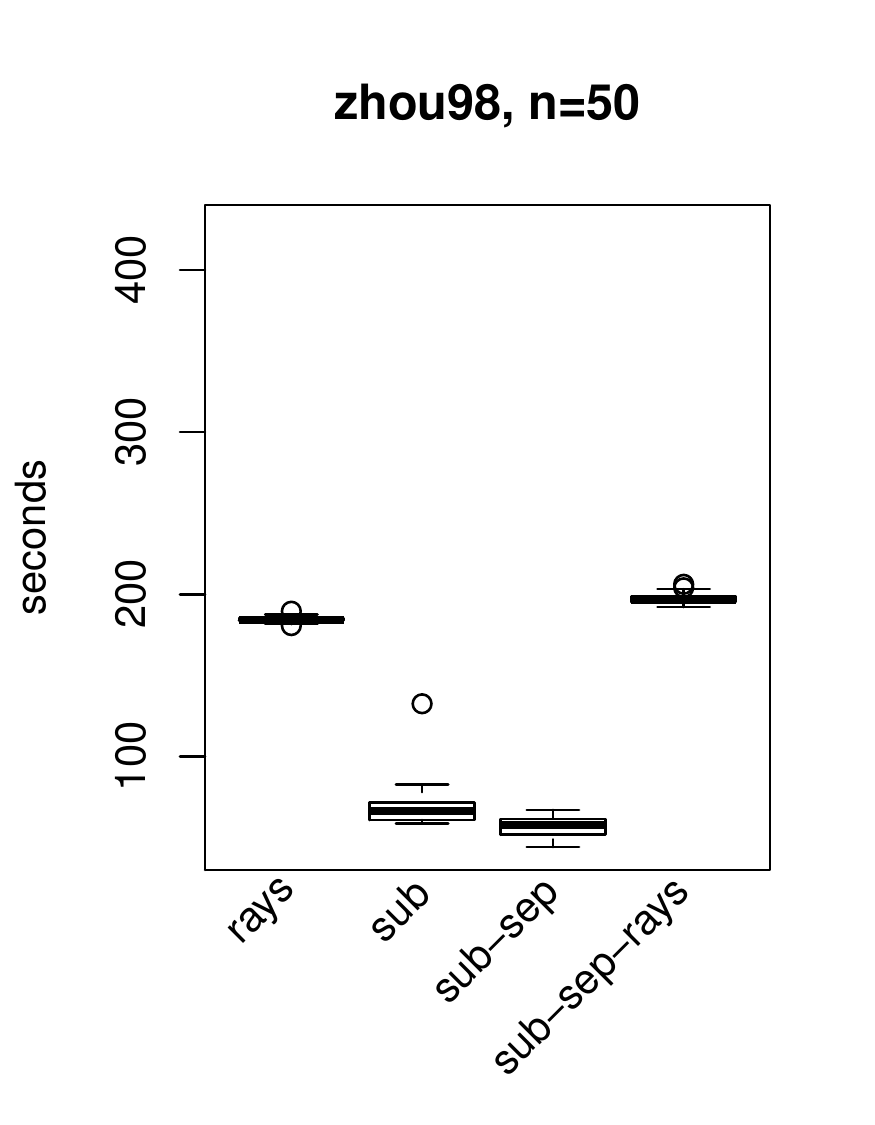}
\includegraphics[scale=0.6,trim=30 72 0 0,clip=TRUE]{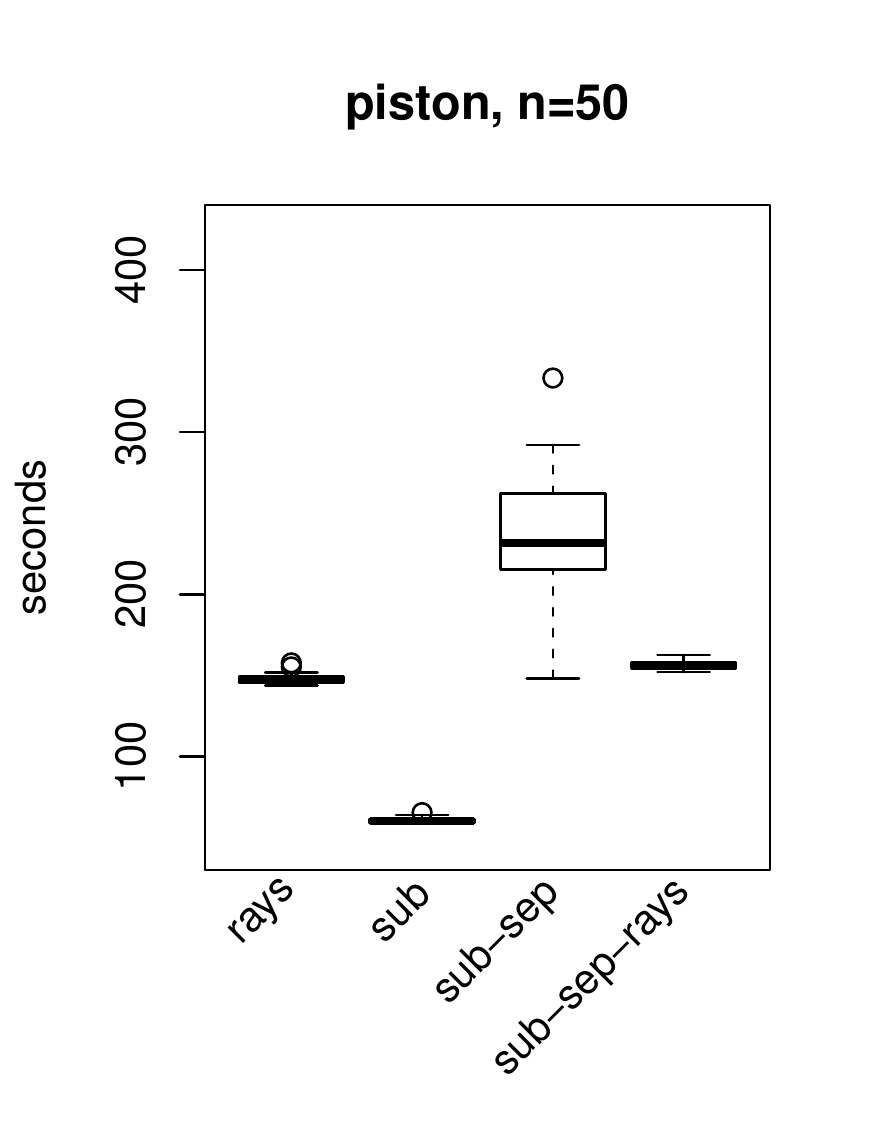} \\
\includegraphics[scale=0.6,trim=0 0 5 50,clip=TRUE]{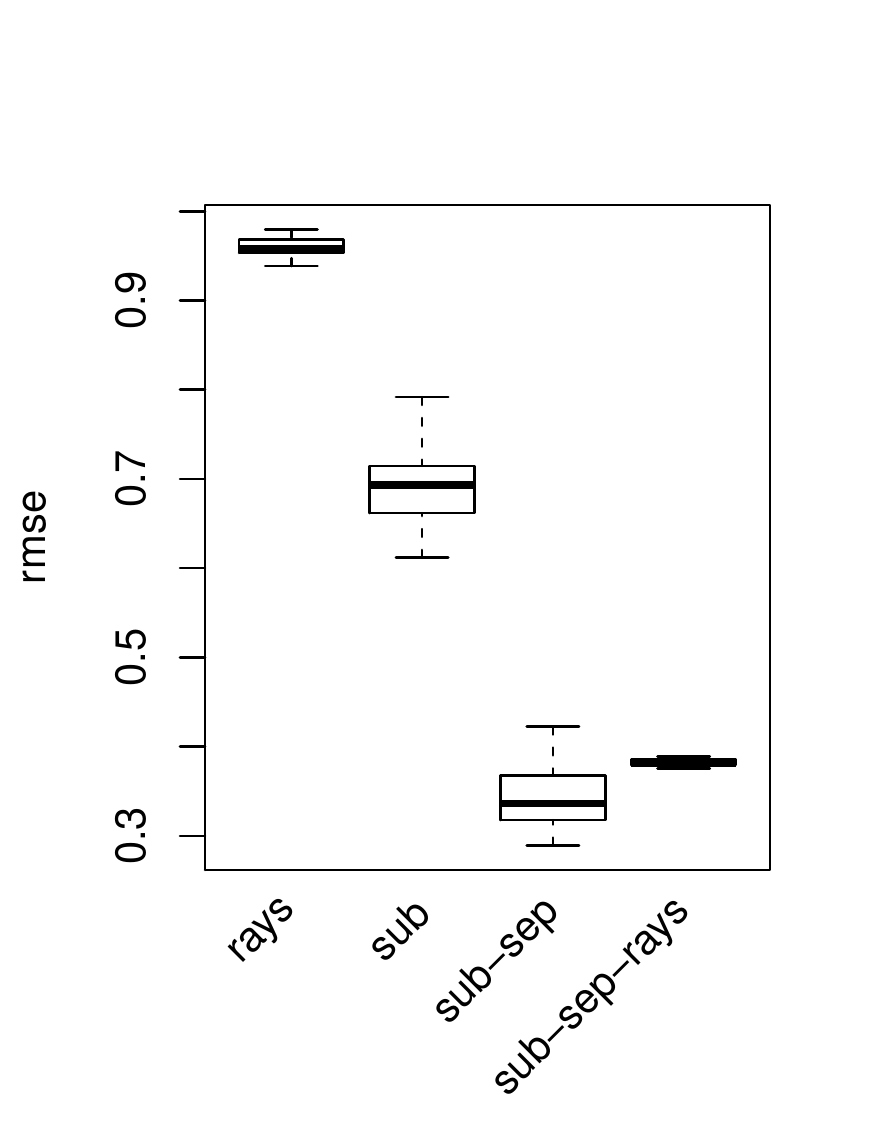}
\includegraphics[scale=0.6,trim=30 0 5 50,clip=TRUE]{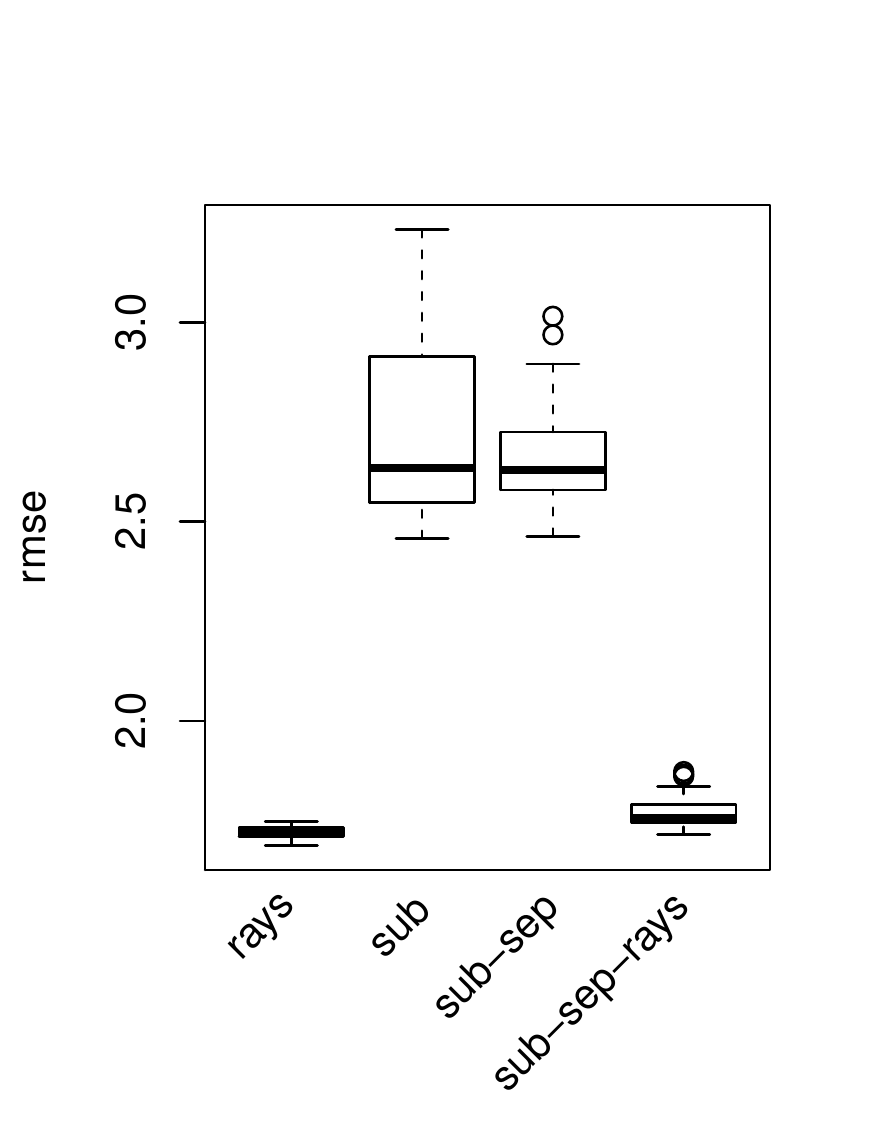}
\includegraphics[scale=0.6,trim=30 0 0 50,clip=TRUE]{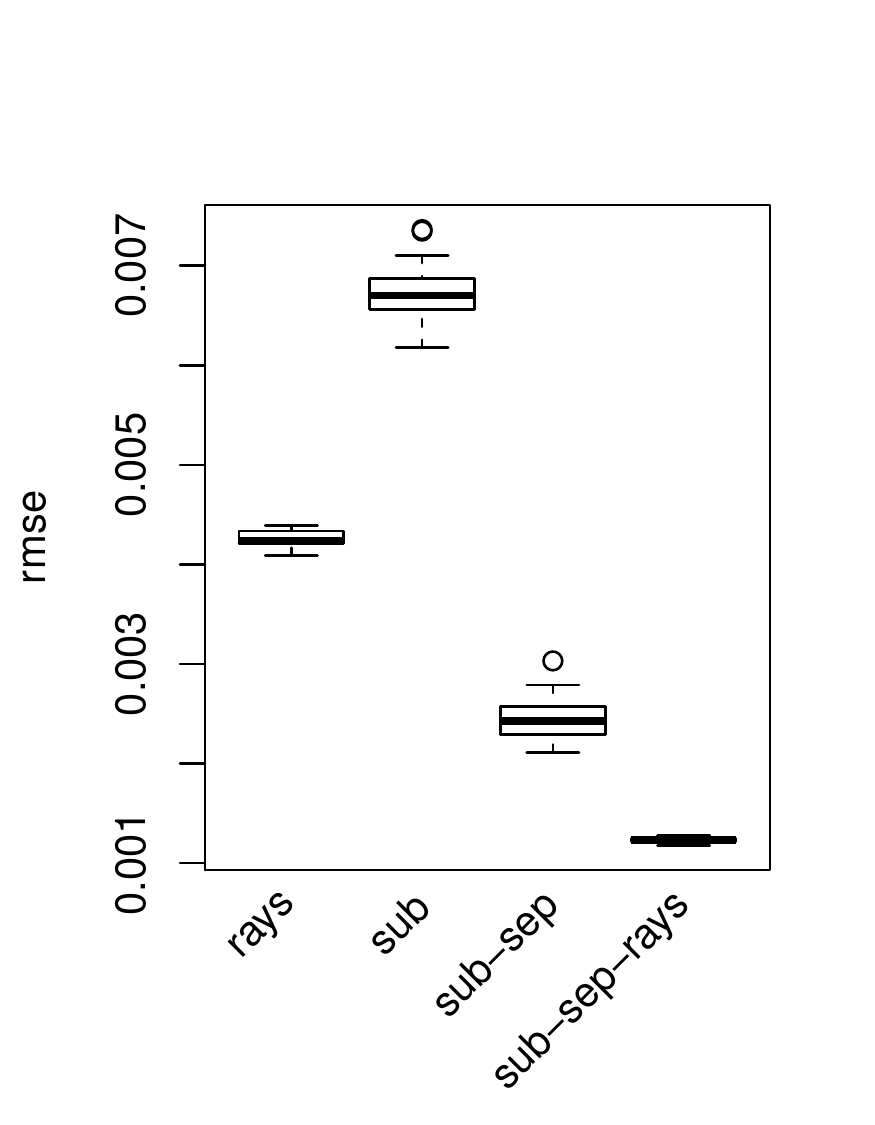}
\caption{Comparing compute time (seconds) and RMSE for ray-based local approximate
GP predictors, and random subset-based GP preditors, for three benchmark
computer experiment problems.}
\label{f:mc}
\end{figure}

Figure \ref{f:mc} shows six boxplots, paired vertically, with a column for
each of three simulation setups.  The top row summarizes compute times in
seconds and bottom one has RMSEs obtained over thirty repeated Monte Carlo
test and training sets.  The borehole experiment used LHS training sets with
$N=64$K, and the other two used $N=40$K.  All three involved LHS testing sets
of size 10000.  The local approximate GP method, via rays, used $n=60$ for the
borehole data, and $n=50$ for the other two.  Large choices of $N$ and $n$ for
the borehole experiment, relative to the others, are motivated by the
smoothness observed in the previous comparison.  Comparisons are drawn to 
random-subsample GP predictors using an isotropic correlation (``sub''), 
a separable version (``sub-sep''), and a rays version applied after re-scaling
the inputs based on (the square root of) those estimated lengthscales.

A result common to all three experiments, echoing earlier observations, is
that the variability in sub-sampled GP estimators, both by time and by RMSE,
is high.  Given a particular random training design, it is hard to predict how
many iterations, involving expensive $1000 \times 1000$ matrix inverses, will
be required to learn the lengthscale(s).  The local approximation(s) perform
best in four of the six comparisons being made to sub-sampled versions with
the same correlation structure.  Estimating a separable global lenghscale is
clearly desirable, whether for local or global emulation, if global isotropy
is a poor approximation.  The Zhou (1998) data is highly isotropic, so
only in that case are the isotropic methods best.

\begin{figure}[ht!]
\centering
\includegraphics[scale=0.57]{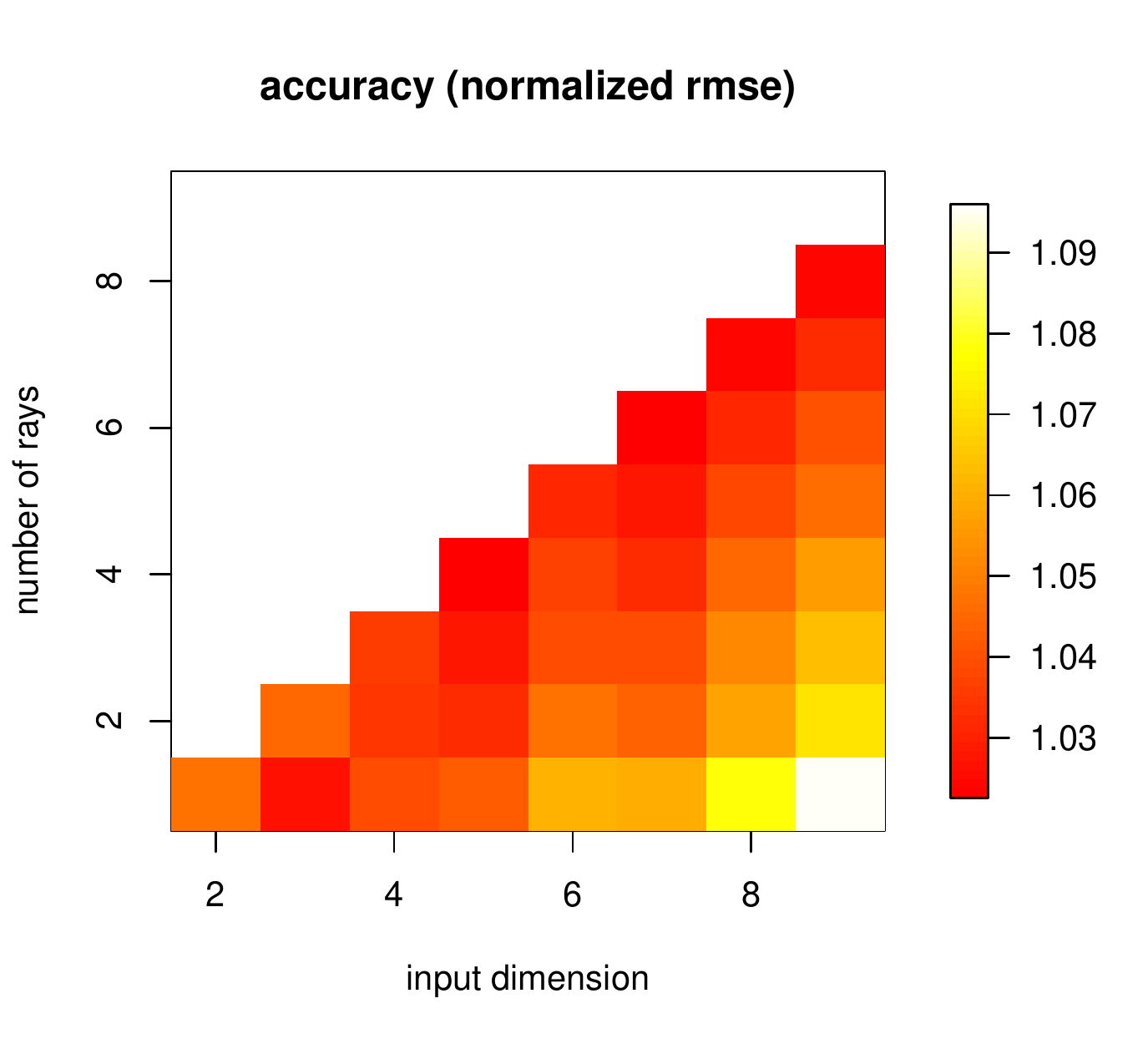} \hspace{0.3cm}
\includegraphics[scale=0.57]{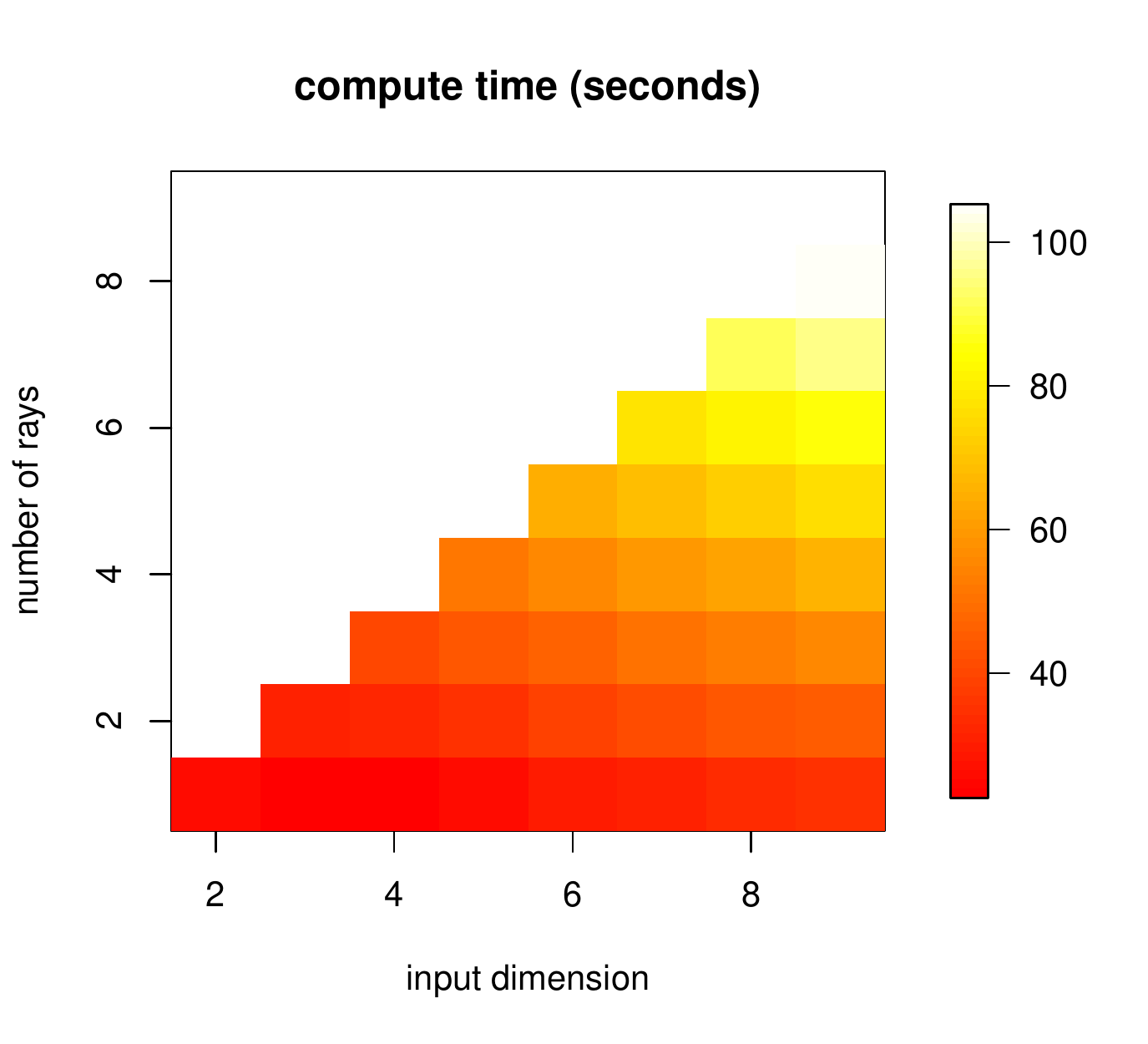}
\caption{RMSE {\em (left)} and time {\em (right)} in predicting
Zhou (1998) data as a function of the input dimension and number of rays
used for local searches.}
\label{f:zhou98nr}
\end{figure}

Figure \ref{f:zhou98nr} shows the outcome of variations on Zhou (1998)
experiments as input dimensions range for $p\in \{2, \dots, 9\}$ and the
number of rays used in local search vary from one to $p$.  The image plots report
averages over thirty repeats with (besides the input dimension) the same setup
as described above.  Since the scale of the response changes as $p$ varies,
the RMSEs reported in the {\em left} pane are normalized by the minimum value
obtained across the thirty repetitions.  Observe that, for a fixed number of
rays, RMSE deterioriates as the dimension increases. Allowing the number of
rays to match $p$ keeps them steady.  The color legend indicates that the
performance gain is never better than 10\% on the normalized RMSE scale. Also,
we remark that there is no gain to having even more than $p$ rays (not shown).
The {\em right} pane shows compute time growing more quickly for the number of
rays than for $p$.

\section{Discussion}
\label{sec:discuss}

Approximate Gaussian process (GP) prediction is becoming essential for
emulating simulation experiment output of the size generated by modern
supercomputers.  In this paper we focused on a local approach to emulation,
building up local sub-designs containing input locations nearby to where
predictions are required.  By empirically studying the topology of surfaces
depicting potential for reduction in predictive variance,   the main objective
behind the local design scheme, we discovered a regularity that
was not being exploited by an exhaustive search.  Instead, we suggested a
continuous 1-d line search along rays emanating from the predictive location
of interest, mimicking patterns of local designs observed in practice.  We then
showed, empirically, how the resulting ray-based designs yielded comparable
out-of-sample predictive performance in a fraction of the time.

We discussed how conclusions from the exploratory analysis in Section
\ref{sec:explore} were independent of the choice of correlation function
$K_\theta(\cdot, \cdot)$, in part justifying our use of the simple isotropic
Gaussian family for our empirical work.  We remarked that that choice only
constrains the process locally.  Global predictive properties may still be
non-isotropic, and even non-stationary if parameterization is allowed to vary
over the input space; see \citet{gramacy:apley:2014} for further discussion.
When the response surfaces are highly anisotropic on a global scale, like with
the borehole data, pre-scaling the inputs based on a crude global analysis can
yield substantial improvements; further discussion is offered by
\citet{gramacy:bingham:etal:2014} and in the {\tt laGP} software vignette
\citep{gramacy:2014}. There are many other situations in which the data might
be better modeled by richer families of $K_\theta(\cdot, \cdot)$, and these
certainly {\em would} impact local designs.  In our supplementary material we
provide trajectories of local designs, like the ones in Figure \ref{f:snap},
for several relevant choices. Qualitatively however, similarities outnumber
differences.  Ribbons and rings are still present in the reduction in variance
surface, and the designs that result are still a mix of NNs and satellite
points.

Qualifications may be required in the context of smoothing noisy data, whether
from stochastic simulation experiments or from spatial/geostatistical data
under measurement error.  In that case the covariance structure must be
augmented with a nugget, which in turn must be estimated.  Even in the
deterministic computer experiments context, authors have argued that
estimating a nugget may lead to improved out-of-sample predictive performance
\citep[e.g.,][]{gra:lee:2012,andrianakis:challenor:2012}.  Neither case has yet been
 adequately explored in the local approximate GP context.  Although
\citet{gramacy:2014} provides an example where locally estimated nuggets
accurately capture the heteroskedastic input--output relationship in a simple
example, it is clear that without further regularization applied to the rate
of change of variance (spatially), both mean and variance surfaces will appear
unstable (spatially), especially win low signal-to-noise contexts and in
higher dimension.  In particular, predictive surfaces will lack a sufficient
aesthetic smoothness so as to be visually ``recognizable'' as (approximate) GP
predictors.

Finally, while searching with rays led to big speedups and accurate
approximate emulation for large designs, such schemes are not always
appropriate. We saw that when designs are small and the dimension
large---meaning that the density of points in the space is low---searching
exhaustively is likely to lead to better out-of-sample prediction and, in the
supercomputing setup, better value for a computational budget. Increasing the
number of rays may help improve on accuracy, although the extra computation
may not be warranted in small-to-moderately sized emulation problems.
We also remark here that another disadvantage, especially in the distributed
supercomputing context, is that ray-based searches can have uncertain times to
convergence.  We have observed that the slowest searches can require twice the
number of objective (reduction in variance) evaluations than the average over
a large set of problems ranging over both $j$ and $x$. The exhaustive search,
by comparison, has a deterministic runtime assuming operating system ``noise''
is low.  Random convergence times can present a load balancing challenge when
trying to make the most out of a supercomputing resource. However, as the
timings in Table \ref{t:borehole} show, the ray-based searches are so fast for
the largest problems that perhaps such a small inefficiency might go
unnoticed.

\subsection*{Acknowledgments}

This work was completed in part with resources provided by the University of
Chicago Research Computing Center.  We are grateful for the valuable comments
of an anonymous referee, which led to substantial improvements in the revised draft.

\bibliography{../approx_gp,../gpu/gpu,rays}

\begin{thebibliography}{39}
\newcommand{\enquote}[1]{``#1''}
\expandafter\ifx\csname natexlab\endcsname\relax\def\natexlab#1{#1}\fi

\bibitem[\protect\citename{An and Owen, }2001]{an:owen:2001}
An, J. and Owen, A. (2001).
\newblock \enquote{Quasi-regression.}
\newblock {\em Journal of Complexity\/}, 17, 4, 588--607.

\bibitem[\protect\citename{Andrianakis and Challenor,
  }2012]{andrianakis:challenor:2012}
Andrianakis, I. and Challenor, P. (2012).
\newblock \enquote{The effect of the nugget on {G}aussian process emulators of
  computer models.}
\newblock {\em Computational Statistics \& Data Analysis\/}, 56, 12,
  4125--4228.

\bibitem[\protect\citename{Ben-Ari and Steinberg, }2007]{benari:steinberg:2007}
Ben-Ari, E. and Steinberg, D. (2007).
\newblock \enquote{Modeling Data from Computer Experiments: and empirical
  comparison of kriging with MARS and projection pursuit regression.}
\newblock {\em Quality Engineering\/}, 19, 4, 327--338.

\bibitem[\protect\citename{Berger et~al., }2001]{berger:deiliveira:sanso:2001}
Berger, J., De~Oliveira, V., and Sanso, B. (2001).
\newblock \enquote{Objective Bayesian Analysis of Spatially Correlated Data.}
\newblock {\em Journal of the American Statistical Association\/}, 96,
  1361--1374.

\bibitem[\protect\citename{Brent, }1973]{brent:1973}
Brent, R. (1973).
\newblock {\em Algorithm for Minimization without Derivatives\/}.
\newblock Englewood Cliffs, N.J.: Prentice--Hall.

\bibitem[\protect\citename{Chen et~al., }2013]{chen:loeppky:sacks:welch}
Chen, H., Loeppky, J., Sacks, J., and Welch, W. (2013).
\newblock \enquote{Analysis Methods for Computer Experiments: How to Assess and
  What Counts?}
\newblock Tech. rep., University of British Columbia.

\bibitem[\protect\citename{Chipman et~al., }2012]{chipman:rajan:wang:2012}
Chipman, H., Ranjan, P., and Wang, W. (2012).
\newblock \enquote{Sequential Design for Computer Experiments with a Flexible
  Bayesian Additive Model.}
\newblock {\em Canadian Journal of Statistics\/}, 40, 4, 663--678.

\bibitem[\protect\citename{Cressie, }1991]{cressie:1993}
Cressie, N. (1991).
\newblock {\em Statistics for Spatial Data, {\em revised edition}\/}.
\newblock John Wiley and Sons, Inc.

\bibitem[\protect\citename{Cressie and Johannesson, }2008]{cressie:joh:2008}
Cressie, N. and Johannesson, G. (2008).
\newblock \enquote{Fixed Rank Kriging for Very Large Data Sets.}
\newblock {\em Journal of the Royal Statistical Soceity, Series B\/}, 70, 1,
  209--226.

\bibitem[\protect\citename{Datta et~al., }2014]{datta:etal:2014}
Datta, A., Banerjee, S., Finley, A.~O., and Gelfand, A.~E. (2014).
\newblock \enquote{Hierarchical Nearest-Neighbor {G}aussian Process Models for
  Large Geostatistical Datasets.}
\newblock Tech. rep., Univeristy of Minnesota.
\newblock Ar{X}iv:1406.7343.

\bibitem[\protect\citename{Eidsvik et~al., }2013]{eidsvik2013estimation}
Eidsvik, J., Shaby, B.~A., Reich, B.~J., Wheeler, M., and Niemi, J. (2013).
\newblock \enquote{Estimation and prediction in spatial models with block
  composite likelihoods.}
\newblock {\em Journal of Computational and Graphical Statistics\/}, 0,
  Accepted for publication, --.

\bibitem[\protect\citename{Emory, }2009]{emory:2009}
Emory, X. (2009).
\newblock \enquote{The kriging update equations and their application to the
  selection of neighboring data.}
\newblock {\em Computational Geosciences\/}, 13, 3, 269--280.

\bibitem[\protect\citename{Franey et~al., }2012]{franey:ranjan:chipman:2012}
Franey, M., Ranjan, P., and Chipman, H. (2012).
\newblock \enquote{A Short Note on {G}aussian Process Modeling for Large
  Datasets using Graphics Processing Units.}
\newblock Tech. rep., Acadia University.

\bibitem[\protect\citename{Gramacy, }2014]{gramacy:2014}
Gramacy, R. (2014).
\newblock \enquote{{\tt laGP}: Large-Scale Spatial Modeling via Local
  Approximate {G}aussian Processes in {\sf R}.}
\newblock Tech. rep., The University of Chicago.
\newblock Available as a vignette in the {\tt laGP} package.

\bibitem[\protect\citename{Gramacy et~al.,
  }2014{\natexlab{a}}]{gramacy:bingham:etal:2014}
Gramacy, R., Bingham, D., Holloway, J.~P., Grosskopf, M.~J., Kuranz, C.~C.,
  Rutter, E., Trantham, M., and Drake, P.~R. (2014{\natexlab{a}}).
\newblock \enquote{Calibrating a large computer experiment simulating radiative
  shock hydrodynamics.}
\newblock Tech. rep., The University of Chicago.
\newblock Ar{X}iv:1410.3293.

\bibitem[\protect\citename{Gramacy and Lee, }2012]{gra:lee:2012}
Gramacy, R. and Lee, H. (2012).
\newblock \enquote{Cases for the nugget in modeling computer experiments.}
\newblock {\em Statistics and Computing\/}, 22, 3.

\bibitem[\protect\citename{Gramacy et~al.,
  }2014{\natexlab{b}}]{gramacy:niemi:weiss:2014}
Gramacy, R., Niemi, J., and Weiss, R. (2014{\natexlab{b}}).
\newblock \enquote{Massively Parallel Approximate Gaussian Process Regression.}
\newblock Tech. rep., The University of Chicago.
\newblock Ar{X}iv:1310.5182.

\bibitem[\protect\citename{Gramacy and Polson, }2011]{gramacy:polson:2011}
Gramacy, R. and Polson, N. (2011).
\newblock \enquote{Particle Learning of {G}aussian Process Models for
  Sequential Design and Optimization.}
\newblock {\em Journal of Computational and Graphical Statistics\/}, 20, 1,
  102--118.

\bibitem[\protect\citename{Gramacy, }2013]{laGP}
Gramacy, R.~B. (2013).
\newblock {\em {\tt laGP}: {L}ocal approximate {G}aussian process
  regression\/}.
\newblock R package version 1.0.

\bibitem[\protect\citename{{Gramacy} and {Apley}, }2014]{gramacy:apley:2014}
{Gramacy}, R.~B. and {Apley}, D.~W. (2014).
\newblock \enquote{Local Gaussian process approximation for large computer
  experiments.}
\newblock {\em Journal of Computational and Graphical Statistics\/}.
\newblock {\em to appear}; see ar{X}iv:1303.0383.

\bibitem[\protect\citename{Gramacy and Lee, }2009]{gra:lee:2009}
Gramacy, R.~B. and Lee, H. K.~H. (2009).
\newblock \enquote{Adaptive Design and Analysis of Supercomputer Experiments.}
\newblock {\em Technometrics\/}, 51, 2, 130--145.

\bibitem[\protect\citename{Gramacy et~al., }2012]{gra:tadd:wild:2012}
Gramacy, R.~B., Taddy, M.~A., and Wild, S. (2012).
\newblock \enquote{Variable Selection and Sensitivity Analysis via Dynamic
  Trees with an Application to Computer Code Performance Tuning.}
\newblock {\em Annals of Applied Statistics\/}, 7, 1, 51--80.
\newblock Ar{X}iv: 1108.4739.

\bibitem[\protect\citename{Haaland and Qian, }2011]{haaland:qian:2012}
Haaland, B. and Qian, P. (2011).
\newblock \enquote{Accurate Emulators for Large-Scale Computer Experiments.}
\newblock {\em Annals of Statistics\/}, 39, 6, 2974--3002.

\bibitem[\protect\citename{Kaufman et~al., }2012]{kaufman:etal:2012}
Kaufman, C., Bingham, D., Habib, S., Heitmann, K., and Frieman, J. (2012).
\newblock \enquote{Efficient Emulators of Computer Experiments Using Compactly
  Supported Correlation Functions, With An Application to Cosmology.}
\newblock {\em Annals of Applied Statistics\/}, 5, 4, 2470--2492.

\bibitem[\protect\citename{Kenett, }1998]{kenett:zacks:1998}
Kenett, R. anmd~Zacks, S. (1998).
\newblock {\em Modern Industrial Statistics: design and control of quality and
  reliability\/}.
\newblock Pacific Grove, CA: Duxbury Press.

\bibitem[\protect\citename{Morris et~al.,
  }1993]{morris:mitchell:ylvisaker:1993}
Morris, D., Mitchell, T., and Ylvisaker, D. (1993).
\newblock \enquote{Bayesian Design and Analysis of Computer Experiments: Use of
  Derivatives in Surface Prediction.}
\newblock {\em Technometrics\/}, 35, 243--255.

\bibitem[\protect\citename{Paciorek et~al., }2013]{paciorek:etal:2013}
Paciorek, C., Lipshitz, B., Zhuo, W., Prabhat, Kaufman, C., and Thomas, R.
  (2013).
\newblock \enquote{Parallelizing {G}aussian Process Calculations in {\sf R}.}
\newblock Tech. rep., University of California, Berkeley.
\newblock Ar{X}iv:1303.0383.

\bibitem[\protect\citename{Pratola et~al., }2014]{pratola:etal:2013}
Pratola, M.~T., Chipman, H., Gattiker, J., Higdon, D., McCulloch, R., and Rust,
  W. (2014).
\newblock \enquote{Parallel Bayesian Additive Regression Trees.}
\newblock {\em Journal of Computational and Graphical Statistics\/}, 23, 3,
  830--852.

\bibitem[\protect\citename{Sacks et~al., }1989]{sack:welc:mitc:wynn:1989}
Sacks, J., Welch, W.~J., Mitchell, T.~J., and Wynn, H.~P. (1989).
\newblock \enquote{Design and Analysis of Computer Experiments.}
\newblock {\em Statistical Science\/}, 4, 409--435.

\bibitem[\protect\citename{Sang and Huang, }2012]{sang:huang:2012}
Sang, H. and Huang, J.~Z. (2012).
\newblock \enquote{A Full Scale Approximation of Covariance Functions for Large
  Spatial Data Sets.}
\newblock {\em Journal of the Royal Statistical Society: Series B\/}, 74, 1,
  111--132.

\bibitem[\protect\citename{Santner et~al., }2003]{sant:will:notz:2003}
Santner, T.~J., Williams, B.~J., and Notz, W.~I. (2003).
\newblock {\em The Design and Analysis of Computer Experiments\/}.
\newblock New York, NY: Springer-Verlag.

\bibitem[\protect\citename{Seo et~al., }2000]{seo:etal:2000}
Seo, S., Wallat, M., Graepel, T., and Obermayer, K. (2000).
\newblock \enquote{Gaussian Process Regression: Active Data Selection and Test
  Point Rejection.}
\newblock In {\em Proceedings of the International Joint Conference on Neural
  Networks\/}, vol. III,  241--246. IEEE.

\bibitem[\protect\citename{Snelson and Ghahramani, }2006]{snelson:ghahr:2006}
Snelson, E. and Ghahramani, Z. (2006).
\newblock \enquote{Sparse Gaussian Processes using Pseudo-inputs.}
\newblock In {\em Advances in Neural Information Processing Systems\/},
  1257--1264. MIT press.

\bibitem[\protect\citename{Stein, }1999]{stein:1999}
Stein, M.~L. (1999).
\newblock {\em Interpolation of Spatial Data\/}.
\newblock New York, NY: Springer.

\bibitem[\protect\citename{Stein et~al., }2004]{stein:chi:welty:2004}
Stein, M.~L., Chi, Z., and Welty, L.~J. (2004).
\newblock \enquote{Approximating Likelihoods for Large Spatial Data Sets.}
\newblock {\em Journal of the Royal Statistical Society, Series B\/}, 66, 2,
  275--296.

\bibitem[\protect\citename{Surjanovic and Bingham, }2014]{simulationlib}
Surjanovic, S. and Bingham, D. (2014).
\newblock \enquote{Virtual Library of Simulation Experiments: Test Functions
  and Datasets.}
\newblock Retrieved December 4, 2014, from \url{http://www.sfu.ca/~ssurjano}.

\bibitem[\protect\citename{Vecchia, }1988]{vecchia:1988}
Vecchia, A. (1988).
\newblock \enquote{Estimation and model identification for continuous spatial
  processes.}
\newblock {\em Journal of the Royal Statistical Soceity, Series B\/}, 50,
  297--312.

\bibitem[\protect\citename{Worley, }1987]{worley:1987}
Worley, B. (1987).
\newblock \enquote{Deterministic Uncertainty Analysis.}
\newblock Tech. Rep. ORN-0628, National Technical Information Service, 5285
  Port Royal Road, Springfield, VA 22161, USA.

\bibitem[\protect\citename{Zhou, }1998]{zhou:1998}
Zhou, Y. (1998).
\newblock \enquote{Adaptive importance sampling for integration.}
\newblock Ph.D. thesis.

\end{thebibliography}
\bibliographystyle{jasa}

\end{document}